\documentclass[
prl,
twocolumn,
superscriptaddress
]{revtex4-1}
\pdfoutput=1
\usepackage[letterpaper,top=1.45cm,bottom=1.65cm,left=2cm,right=2cm]{geometry}
\usepackage{amssymb,amsmath,amsthm,graphicx,mathptmx}
\usepackage{subfigure}
\usepackage{graphics, color}
\usepackage{latexsym}
\usepackage{bm}
\usepackage{epsfig}
\usepackage{multirow,tabularx}
\usepackage[colorlinks=true,linktocpage=true,linkcolor=blue,citecolor=blue]{hyperref}
\usepackage[abs]{overpic}

\usepackage{fancyhdr}
\fancypagestyle{plain}

\newcommand{\PRLsection}[1]{\emph{#1.---}}

\newcolumntype{Y}{>{\centering\arraybackslash}X}

\begin{document}
\title{End Point of the Ultraspinning Instability and Violation of Cosmic Censorship}
\author{Pau Figueras}%
\email{p.figueras@qmul.ac.uk}
\affiliation{School of Mathematical Sciences, Queen Mary University of London, Mile End Road, London E1 4NS, United Kingdom}%
\author{Markus Kunesch}%
\email{m.kunesch@damtp.cam.ac.uk}
\affiliation{Department of Applied Mathematics and Theoretical Physics (DAMTP), Centre for Mathematical Sciences, University of Cambridge, Wilberforce Road, Cambridge CB3 0WA, United Kingdom}
\author{Luis Lehner}
\email{llehner@perimeterinstitute.ca}
\affiliation{Perimeter Institute for Theoretical Physics, Waterloo, Ontario N2L 2Y5, Canada}
\author{Saran Tunyasuvunakool}
\altaffiliation[now at ]{DeepMind}
\email{stun@google.com}
\affiliation{Department of Applied Mathematics and Theoretical Physics (DAMTP), Centre for Mathematical Sciences, University of Cambridge, Wilberforce Road, Cambridge CB3 0WA, United Kingdom}


\begin{abstract}
We determine the end point of the axisymmetric ultraspinning instability of asymptotically flat Myers-Perry black holes in $D=6$ spacetime dimensions. In the non-linear regime, this instability gives rise to a sequence of
concentric rings connected by segments of black membrane on the rotation plane. The latter become thinner over time,
resulting in the formation of a naked singularity in finite asymptotic time and hence a violation of the weak cosmic
censorship conjecture in asymptotically flat higher-dimensional spaces.
\end{abstract}


\maketitle
\thispagestyle{fancy}

\PRLsection{Introduction}%
The recent detection of gravitational waves from black hole binary mergers \cite{Abbott:2016blz,Abbott:2016nmj} has
provided the first direct observation of these objects. The current observational data are compatible with the predictions
of general relativity, and they suggest that the end point of such mergers is a Kerr black hole (BH) \cite{Kerr:1963ud}. These
observations provide evidence that the Kerr BH in vacuum is non-linearly stable, at least within a certain range of the
angular momentum. However, a mathematically rigorous understanding of the stability of the generic Kerr BH, as well as a thorough understanding of its dynamics under arbitrary perturbations, is still lacking. 
In fact, recent work suggests that novel and nontrivial dynamics may be present very 
close to extremality (e.g.,~\cite{Aretakis:2013dpa,Yang:2014tla,Gralla:2016sxp}).

Higher dimensional BHs, however, can be unstable under gravitational perturbations. This was first shown by
Gregory and Laflamme (GL) for black strings and black $p$-branes \cite{Gregory:1993vy}. Determining the end point of this
instability has been a subject of intense study due to the potential implications on the weak cosmic censorship
conjecture (WCC) in such spacetimes.
With the aid of numerical relativity (NR), \cite{Lehner:2010pn} found that the GL instability gives rise to a
self-similar structure of bulges connected by ever thinner string segments, which all undergo the GL instability.
Eventually, the black string pinches off in finite asymptotic time, resulting in a naked singularity.
Since no fine-tuning of the initial data was required, this result constituted a violation of the WCC, albeit in spacetimes with compact
extra dimensions. 

Contrary to the $D=4$ case, asymptotically flat BHs in higher dimensions can carry arbitrarily large angular
momenta. At very large angular momenta, BHs become highly deformed and resemble black branes, which are known to be
unstable under the GL instability~\cite{Emparan:2003sy}. This observation highlighted the possibility that higher
dimensional asymptotically flat BHs can be unstable under gravitational perturbations. This indeed turned out to be the
case. For instance, the black rings of \cite{Emparan:2001wn} suffer from various types of instabilities
\cite{Elvang:2006dd,Figueras:2011he,Santos:2015iua,Tanabe:2015hda,Figueras:2015hkb,Tanabe:2016pjr}, including the GL
instability. The non-linear evolution of the latter was studied in a very recent work
by three of us~\cite{Figueras:2015hkb}, where
it was found that, for sufficiently thin rings, the evolution of the instability is similar to that of the GL
instability of black strings. Hence, a naked singularity should form in finite asymptotic time, thus violating the WCC in higher-dimensional asymptotically flat spaces. However, the calculations in
\cite{Figueras:2015hkb} were computationally highly demanding, which limited the extent to which the instability could be explored. 
It was therefore not possible to estimate the timescale of a possible pinch-off
or to determine whether the process is self-similar as for black strings.

Ref. \cite{Emparan:2003sy} conjectured (and \cite{Dias:2009iu} later confirmed) that rapidly spinning Myers-Perry (MP)
BHs \cite{Myers:1986un}\footnote{MP BHs are the higher-dimensional analogues of Kerr BHs.} in $D\geq 6$ are unstable under a GL-type
of instability, which is referred to as the ``ultraspinning instability''.  
As with the GL case, there exist zero modes that
connect MP BHs with different families of ``bumpy'' BHs \cite{Dias:2014cia,Emparan:2014pra}.  In this
Letter, we report on the final stages of the evolution of the ultraspinning instability of singly spinning MP BHs in 6
dimensions. We restrict ourselves to the instability that deforms the horizon without breaking any of the rotational symmetries of the
background, i.e., the axisymmetric one. The imposed symmetries reduce the problem to a system of
$(2+1)$-dimensional PDEs, which is significantly more computationally tractable than the 
one described in~\cite{Figueras:2015hkb}. This
allows us to elucidate the dynamics of the ultraspinning instability in full detail.

\PRLsection{Numerical methods}%
We solve the $D=6$ vacuum Einstein equation numerically with a $U(1)\times SO(3)$ isometry imposed. The
ultraspinning instability lies within this symmetry sector, but other non-axisymmetric instabilities which
are not captured by this ansatz do exist. We impose the symmetry using the modified cartoon method \cite{Pretorius:2004jg,Shibata:2010wz,Cook:2016soy}. We employ the CCZ4 formulation \cite{Alic:2011gg,Weyhausen:2011cg} on a
Cartesian grid with the redefinition of the constraint damping parameter $\kappa_1 \to \kappa_1/\alpha$, where $\alpha$ is the
lapse function \cite{Alic:2013xsa}.  Typically, we choose $\kappa_1 = 0.5$ and $\kappa_2=0$. As initial data, we take the 6-dimensional singly spinning MP BH,
\begin{equation}
\begin{aligned}
ds^2 =& -dt^2 + \frac{\mu}{r\,\Sigma}(dt - a\,\sin^2\theta\,d\phi)^2 + \frac{\Sigma}{\Delta}\,dr^2 + \Sigma\,d\theta^2 \\
&+ (r^2 + a^2)\,\sin^2\theta\,d\phi^2 + r^2\,\cos^2\theta\,d\Omega_{(2)}^2\,,\\
\end{aligned}
\label{eq:MPmetric}
\end{equation}
with a new quasi-radial coordinate $\rho$ defined by
\begin{align}
r =~\rho\left(1+\frac{1}{4}\,\frac{r_h^3}{\rho^3}\right)^{\frac{2}{3}}\,,
\end{align}
where $\mu$ and $a$ are the mass and rotation parameters respectively, $\Sigma = r^2 + a^2\,\cos^2\theta$,
$\Delta = r^2 + a^2 - \mu/r$, and $r_h$ is the largest real root of $\Delta(r_h) = 0$.
In our simulations, we set $\mu=1$ and
consider MP BHs with $1.5\leq a/\mu^\frac{1}{3} \leq 2.0$.
The first (ring-shaped) unstable mode sets in at
$a/\mu^\frac{1}{3}=1.572$ and the second (saturn-shaped) mode sets in at $a/\mu^\frac{1}{3}=1.849$ \cite{Dias:2009iu}.

\begin{figure}
\flushright
\begin{overpic}[width=0.97\columnwidth]{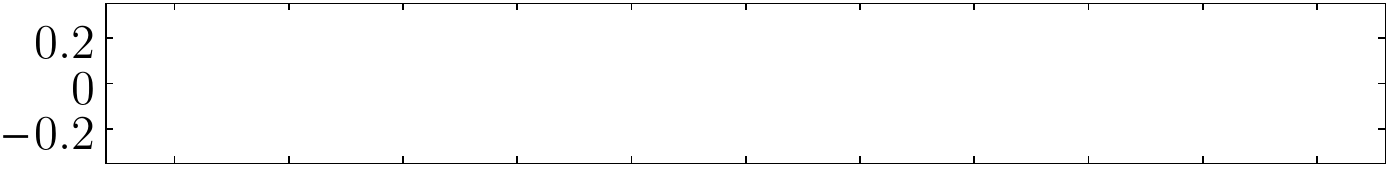}
\put(-2,0){ \includegraphics[width=0.97\columnwidth]{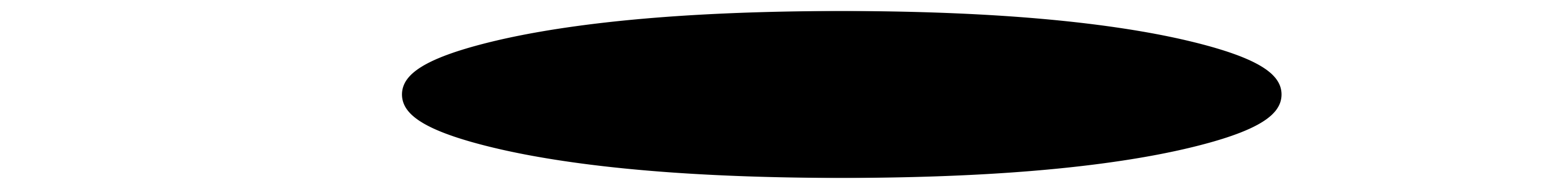}}
\put(22,18){$\hat t=0$}
\end{overpic}
\begin{overpic}[width=0.97\columnwidth]{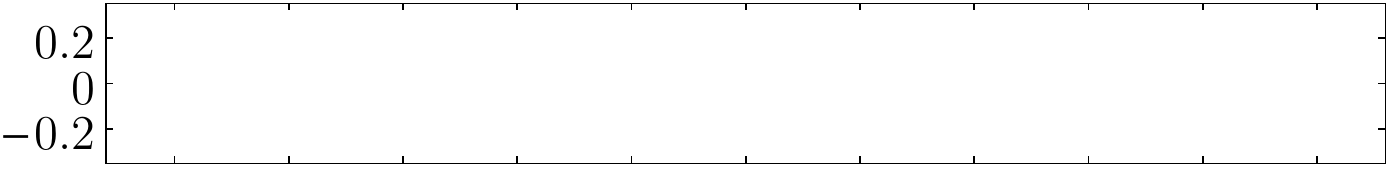}
\put(-2,0){ \includegraphics[width=0.97\columnwidth]{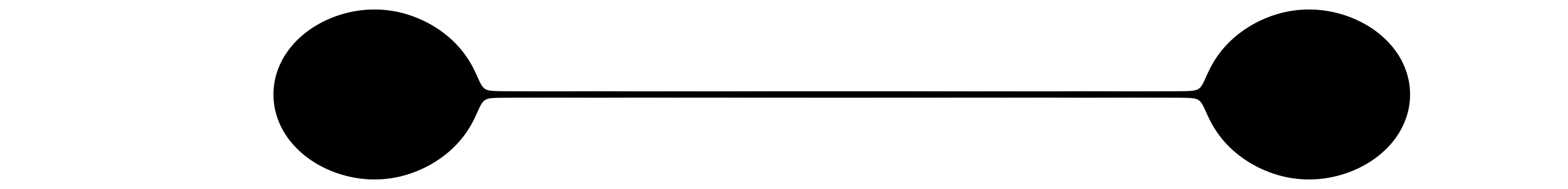}}
\put(107,19){$\hat t=35.857$}
\end{overpic}
\begin{overpic}[width=0.97\columnwidth]{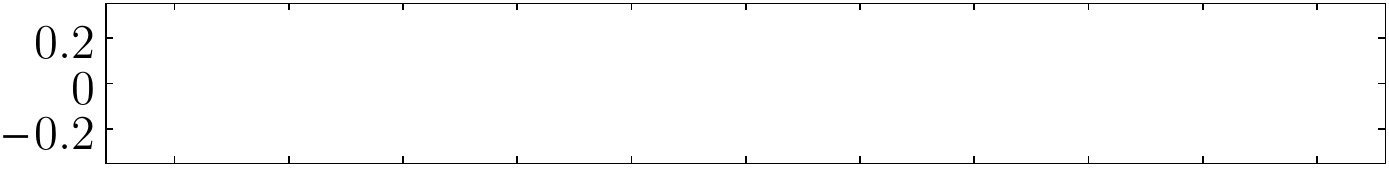}
\put(-2,0){ \includegraphics[width=0.97\columnwidth]{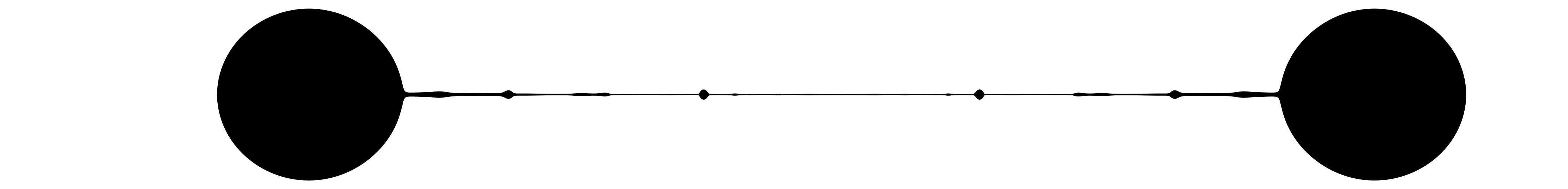}}
\put(-7,26){\rotatebox{90}{$Z$}}
\put(107,19){$\hat t=36.802$}
\end{overpic}
\begin{overpic}[width=0.97\columnwidth]{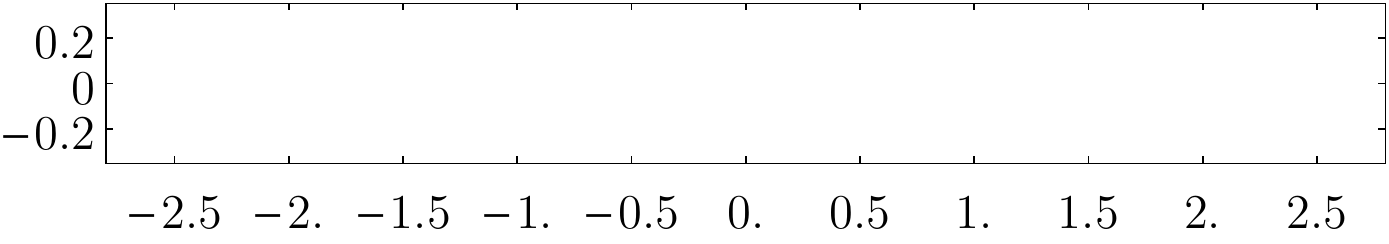}
\put(-2,0){ \includegraphics[width=0.97\columnwidth]{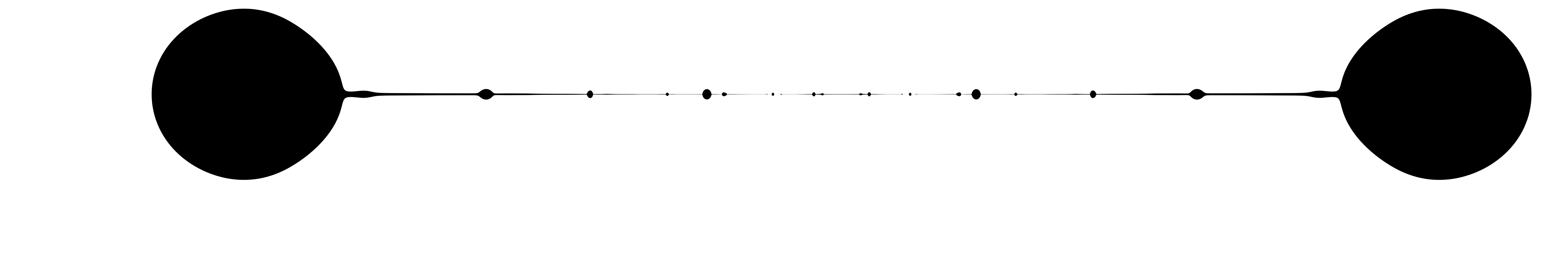}}
\put(107,31){$\hat t=36.9535$}
\end{overpic}
\par
\vspace{0.1cm}
\begin{overpic}[width=0.97\columnwidth]{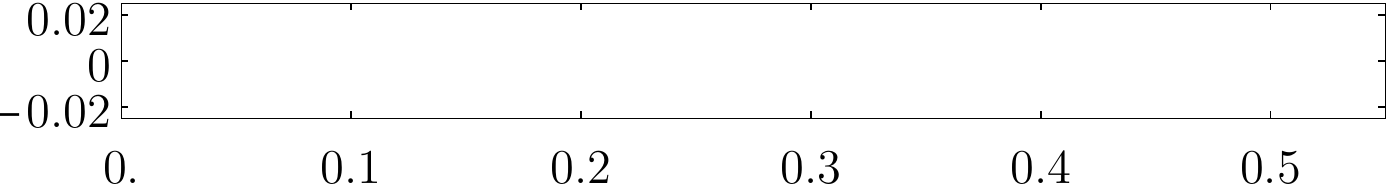}
\put(-2.2,0.5){ \includegraphics[width=0.97\columnwidth]{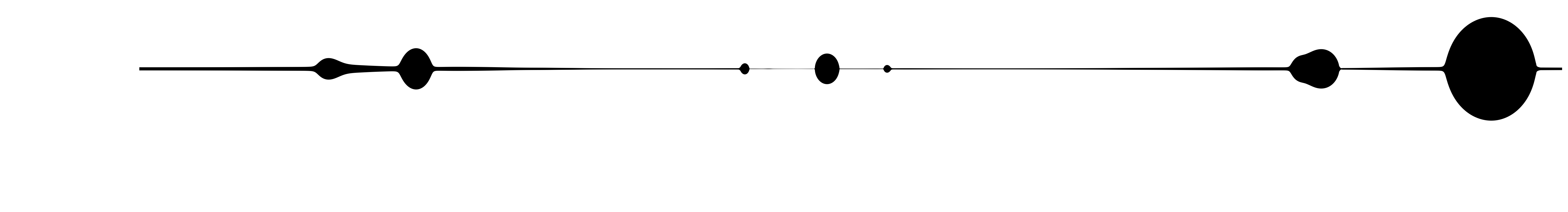}}
\put(110,-4){$U$}
\end{overpic}
\par
\vspace{0.3cm}
\centering
\begin{overpic}[width=\columnwidth]{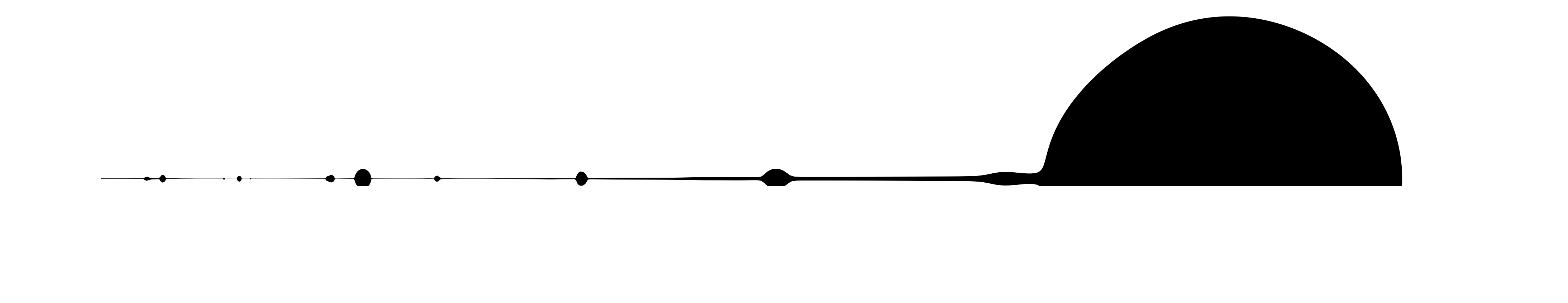}
\put(-2.2,-0.6){ \includegraphics[width=\columnwidth]{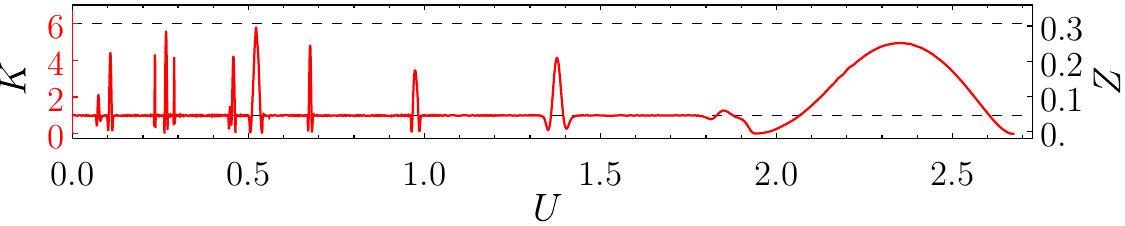}}
\end{overpic}
\caption{\textit{Top}: Embedding diagrams of the AH at different stages of the evolution of the
ultraspinning instability of a MP BH with $a/\mu^\frac{1}{3} = 1.7$.
Here $\hat t = t/\mu^\frac{1}{3}$.
The structure of rings that form on the membrane in the middle depends sensitively on the grid setting
and is not convergent \cite{Supplemental}.
\textit{Bottom}: Normalized spacetime Kretschmann invariant pulled back onto the AH. }
\label{fig:evolutionAH}
\end{figure}

We evolve the lapse and the shift  using the CCZ4 $(1+\log)$
slicing with an advection term and the variant of the Gamma-Driver shift condition used in
\cite{Figueras:2015hkb} (see also the Supplemental Material \cite{Supplemental}).
Initially, we choose $\alpha = \chi$ and $\beta^i = \chi\,\beta^i_\textrm{MP}$, where
$\beta^i_\textrm{MP}$ is the analytic shift obtained from \eqref{eq:MPmetric} and $\chi$ denotes the conformal factor.
To help stabilize the evolution, we add diffusion terms well inside the apparent horizon (AH) as described in \cite{Figueras:2015hkb}. The coordinate singularity present in our initial data is
regularized by the ``turduckening" method \cite{Brown:2007pg,Brown:2008sb}.
Since the gauge in \eqref{eq:MPmetric} is not optimal, we first evolve this
initial data until the gauge has settled to spatial harmonic coordinates with respect to the conformal
metric. In this new gauge, the shape of the AH flattens and resembles a pancake for
rapidly spinning BHs (see Fig. \ref{fig:evolutionAH}), as one would expect on physical grounds
\cite{Emparan:2003sy}. We stress that this gauge adjustment process occurs over a short time period, during which we have verified that there is no significant physical evolution.

Once the gauge dynamics has settled, we trigger the ultraspinning instability by perturbing the conformal factor via
\begin{equation}
\chi = \chi_0\left\{1+A\,J_0 \! \left[j_{0,k}\,\sin \! \left(\tfrac{\pi}{2} \sigma \right) \right]\,\exp \! \left[-\!\left(\tfrac{\chi_0}{\chi_h} -
\tfrac{\chi_h}{\chi_0}\right)^{\! 2} \right]\right\}\!,
\label{eq:perturbation}
\end{equation}
where $A$ is the amplitude of the perturbation,  $\chi_0$ is the unperturbed conformal factor, $\chi_h$ is the value of
the unperturbed conformal factor at the horizon, $J_0$ is the Bessel function of the first kind, $j_{0,k}$ is the
$k^{\textrm{th}}$ zero of $J_0$, and $\sigma = \sqrt{x^2 + y^2}/\tilde R$.
Here $\tilde R$ is a parameter that determines the extent of the deformation in the rotation plane, and $x$ and $y$ are our
Cartesian coordinates. The expression \eqref{eq:perturbation} ensures that the perturbation is localized
on the horizon and behaves like $J_0$ near the rotation axis, where $J_0$ captures the unstable mode reasonably
accurately \cite{Emparan:2003sy}. This perturbation introduces constraint violations,
but they are small and depend linearly on the amplitude of the perturbation.
In our simulations, we check that these constraint violations decay exponentially with time (thanks to the CCZ4 constraint damping terms), and that the physical parameters of the perturbed BH change by less than $1\%$ compared to those of the unperturbed BH.

To understand the end point of the ultraspinning instability, we monitor the geometry of the AH.
Most traditional approaches in NR assume that the AH can be given by the
level set of a function of the angular coordinates (see \cite{Thornburg:2006zb} for a review). In our current symmetry setting
this would mean $r=R(\theta)$, where $r$ is the radial coordinate and $\theta$ is the polar angle on the sphere.
However, in the final stages of the ultraspinning instability, this is not a valid assumption as $R(\theta)$
fails to be a single-valued function (see Fig.  \ref{fig:evolutionAH}).
To overcome this problem, we consider the AH as a completely general parametric surface
$(x(u),y(u),z(u))$, where $u$ is the parameter. We then solve the elliptic PDEs that arise from setting the expansion
and a gauge condition for $u$ to zero.
The technical details of this construction can be found in the Supplemental Material \cite{Supplemental} and in \cite{SaranThesis}.

We solve the CCZ4 equations numerically with the \texttt{GRChombo} code \cite{Clough:2015sqa,chombo-design-doc} using up to $22$ levels of refinement (each refined in a 2:1 ratio) with a coarsest grid spacing of $0.35 \mu^\frac{1}{3}$. We discretize the equations using fourth-order finite differences and integrate in time using RK4. 
We choose our refinement levels such that the AH is covered by at least 57 points at all times.
Convergence studies indicate that the order of convergence is $\approx 3$. Some relevant numerical tests are presented in the Supplemental Material \cite{Supplemental}.

\PRLsection{Results}%
In Fig. \ref{fig:evolutionAH} \textit{Top} we present different snapshots of the embedding of a constant rotational angle section of
the AH into $\mathbb R^4$  at representative stages of the evolution
\footnote{Videos can be found at \url{http://grchombo.github.io}.}.
In the range of $a/\mu^\frac{1}{3}$ that we have explored, the ring mode grows fastest and governs the non-linear evolution.  We find that initially only a large ring forms at the outermost edge of the horizon (second snapshot in Fig. \ref{fig:evolutionAH} \textit{Top}),
even if we perturb with a ``saturn-shaped'' perturbation by setting $k=2$ in \eqref{eq:perturbation}.
In Table \ref{tab:growthRates}, we summarize the growth rates of the first unstable mode for different values of $a/\mu^\frac{1}{3}$
as calculated from our simulations. To our knowledge, these are not currently available in the literature.

\begin{table}[h]
\centering
\begin{tabular}{c|c|c|c|c|c}
\hline
\hline
$a/\mu^{\frac{1}{3}}$ &  1.6 & 1.7 & 1.8 & 1.9 & 2.0 \\
$\Im\varpi\,\mu^\frac{1}{3}$ & $0.020$ & $0.130$ & $0.213$ & $0.262$ & $0.299$ \\
\hline
\hline
\end{tabular}
\caption{Growth rates of the first unstable mode. Errors are $\pm 3\%$ for $a/\mu^{\frac{1}{3}} \geq 1.7$ and $\pm 25\%$
for $a/\mu^\frac{1}{3} = 1.6$.}
\label{tab:growthRates}
\end{table}%

To estimate how much mass and angular momentum are contained within the outermost ring, we calculate the corresponding
Komar integrals on the AH. The calculated mass is only accurate once the system has settled down to a steady
state. Towards the end of our simulations, the Komar mass changes by less than $1\%$, thus indicating that the majority of the AH has settled down sufficiently. We find that the outermost ring accounts for 98--99\% of the total mass and more than 99.99\% of the angular momentum. The radiated mass is too small ($<2\%$) to be distinguishable from changes in the Komar mass due to the system not having settled down completely. Angular momentum is conserved because of our symmetry assumptions.

After the outermost large ring has formed, the region of the horizon connecting it to the rotation axis resembles a thin,
locally boosted black membrane (see Fig. \ref{fig:evolutionAH}). The evolution of the ultraspinning instability takes
place in the radial direction, while the local boost is along an orthogonal $U(1)$ direction. Therefore,
the dynamics of the black membrane under this instability should be insensitive to the local boost and, since the
transverse direction is flat, similar to the GL dynamics of a 5D black string. The portion of the AH that resembles a
black membrane is indeed GL unstable and can accommodate many unstable modes (see Table \ref{tab:generations}).
Its subsequent evolution leads to a sequence of ever thinner rings connected by
segments of black membrane which are GL unstable.

As evidence that the horizon has the geometry of concentric rings connected by membrane sections, we evaluate
the (suitably normalized) spacetime Kretschmann invariant, $K =R_{abcd}R^{abcd}\,Z_\textrm{AH}^4/12$, on the AH.
Here $Z_\textrm{AH}$ is the radius of the transverse sphere, which measures the thickness of
the AH. The normalization is such that $K=1$ for a black membrane and $K=6$ for a 6D black string.
The results (Fig \ref{fig:evolutionAH} \textit{Bottom}) are in close agreement with $K=1$ on the membrane sections
and approach $K=6$ on most of the fully formed rings, suggesting that they are well approximated by stationary black strings.

There are three fundamental differences between the dynamics of unstable black strings and ultraspinning MP BHs. Firstly, the latter have compact horizons that do not wrap any topological direction in spacetime.
Thus, any self-similarity is broken near the edges, and also in the early stages of the instability when the
radial extent of the unstable membrane sections is comparable to the size of the whole BH.
Secondly, MP BHs are rotating and the imposition of axisymmetry introduces a new constraint: the conservation of
angular momentum. Furthermore, the rotation causes a centrifugal force which redistributes angular momentum outwards and leads to different membrane sections having different thicknesses (see third snapshot in Fig. \ref{fig:evolutionAH}).
Hence, the local GL instabilities of each membrane section evolve on different timescales.
Thirdly, throughout the whole evolution, the small concentric rings that form after the first generation move around,
causing significant additional stretching over the time it takes to form a new generation. The non-zero boost velocity imparted upon the membrane delays the formation of the $i^{\text{th}}$ \cite{Hovdebo:2006jy}
generation, while the stretching itself also causes the membrane to become thinner,
resulting in the earlier formation of the $(i+1)^{\text{th}}$ generation.

The combination of these three effects implies that the evolution of the ultraspinning
instability is not self-similar: while we do observe newly formed membrane sections all undergoing the GL
instability, the time elapsed between the formation of successive generations does not decrease with a universal
factor (c.f. Table \ref{tab:generations}), even for later generations. Instead, in the $a/\mu^\frac{1}{3} =1.7$ run,
we observe factors between 0.07 and 0.42.
Furthermore, they cause the pinch-off to happen sooner, mostly due to the quick drop in the formation times between generations at the beginning.
The largest factor between generations that we observed was $X_{\max}=0.41$.
Since $X_{\max}<1$, we can bound the pinch-off time  by a geometric series
\begin{align}
t_c < t_0 + (t_1 - t_0) \sum_i X^{i} < t_0 + (t_1-t_0)/(1-X_{\max}).
\end{align}
While this upper bound is not sharp, it provides evidence that the BH pinches off in finite asymptotic time.

From Table \ref{tab:generations}, we see that the typical ratio $R_i/L_i$ between the thickness and the length of a
membrane section varies between 300 and 600. For the GL instability of black strings, this ratio is approximately 100 across generations \cite{Lehner:2010pn}. Therefore, the membranes that form in the evolution of the ultraspinning instability are more unstable, indicating a faster pinch-off time.

\begin{table}[h]
\centering
\begin{tabular}{c|c|c|c|c|c}
\hline
\hline
Gen. & 1 & 2 & 3 & 4 & 5 \\
\hline
$t_i/\mu^\frac{1}{3}$  & 31.8 & 36.45 & 36.78 & 36.916 & 36.952 \\
$L_i/Z_{AH,i}$  & 540 & 530 & 370 & 510 & $> 370$ \\
\hline
\hline
\end{tabular}
\caption{Properties of the generations. The ratio of length to thickness of the $i^\text{th}$ generation membrane was measured
just before the formation of the $(i+1)^\text{th}$ generation. The time it takes to form the next generation
decreases with factors $0.07$, $0.41$ and $0.26$.}
\label{tab:generations}
\end{table}%

\begin{figure}
\centering
\begin{overpic}[width=\columnwidth]{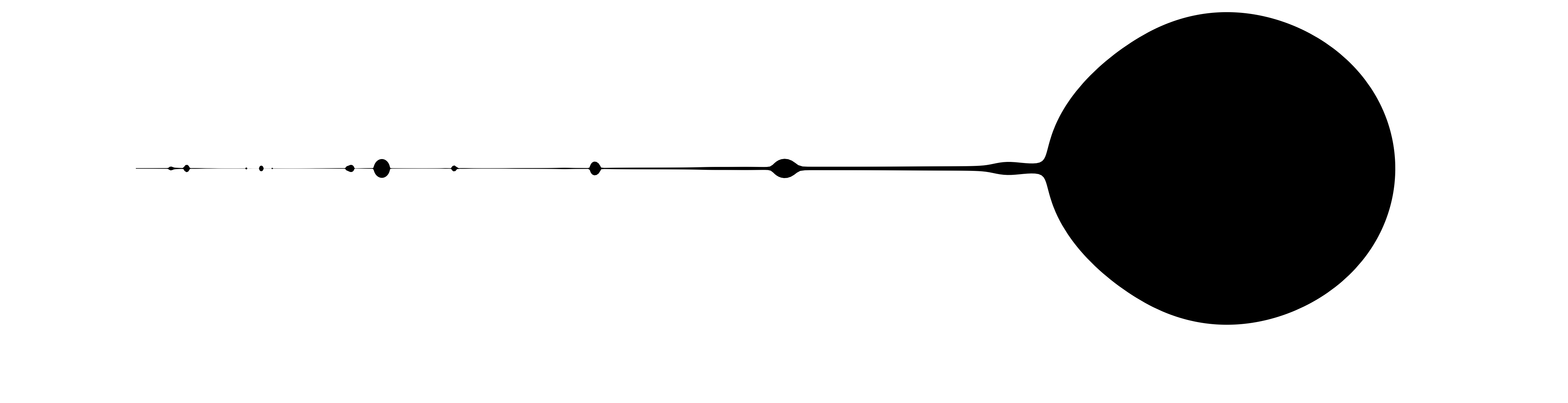}
\put(-2.2,-1.05){ \includegraphics[width=\columnwidth]{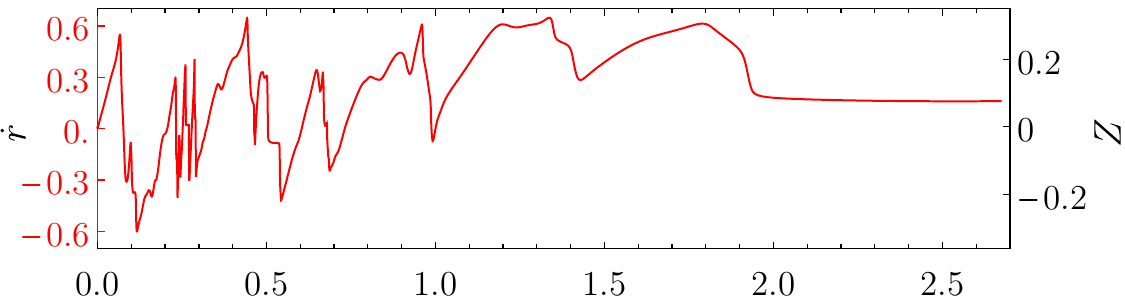}}
\end{overpic}
\par
\vspace*{0.1cm}
\begin{overpic}[width=\columnwidth]{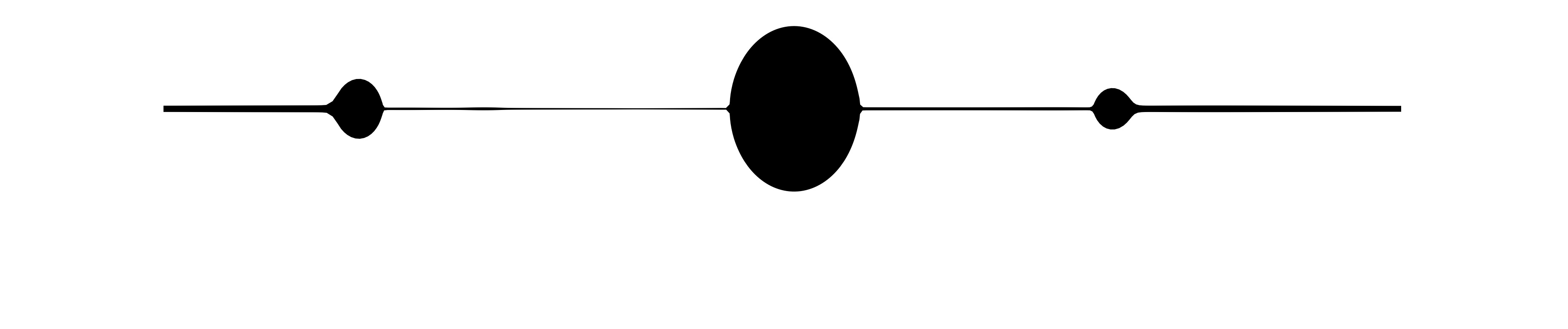}
\put(-2.2,-0.7){ \includegraphics[width=\columnwidth]{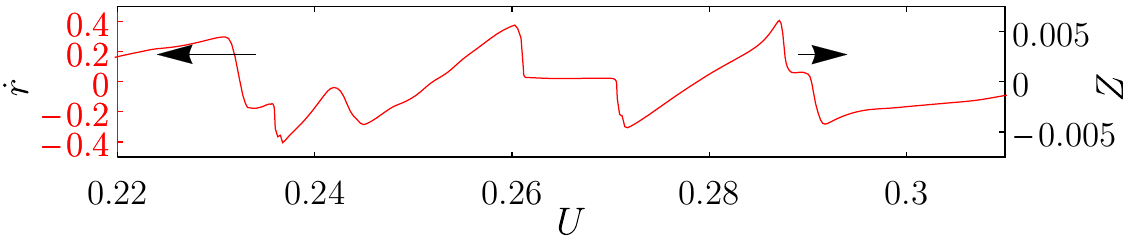}}
\end{overpic}
\par
\vspace*{0.2cm}
\includegraphics[width=0.9\columnwidth]{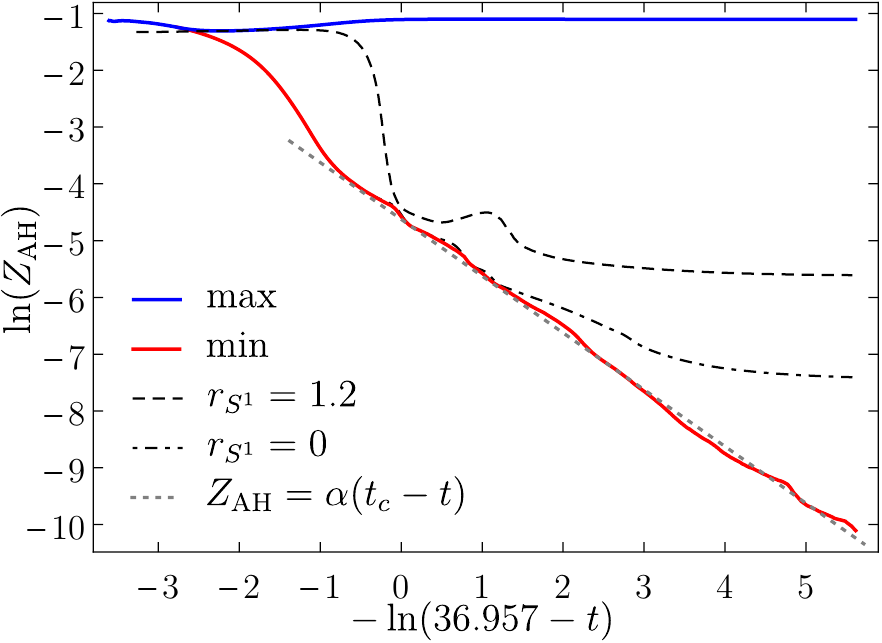}
\caption{\textit{Top}: radial velocity $\dot r\equiv dr/dt$ of a null ray moving with the AH.
\textit{Middle:} Zoom of the thinnest region. The arrows indicate the direction of the local velocity of the rings.
\textit{Bottom}: Evolution of the AH thickness at several representative locations for
$a/\mu^\frac{1}{3}=1.7$. }
\label{fig:thicknessPlot}
\end{figure}

Let us now explain the local dynamics which leads to the non-constant factors between generations. We calculate the radial velocity $dr/dt$ of null rays which co-rotate with the BH to estimate the local radial flow velocity of the AH.
The results are shown in Fig. \ref{fig:thicknessPlot}. They paint a very consistent picture: near each ring, the radial velocity either decreases or reverses completely, leading to a build up of mass. This also explains the numerous sign changes around the thinnest
point of the membrane, where many higher generation rings are present.

The outermost ring very quickly settles down to an almost stationary state.  However, as Fig. \ref{fig:thicknessPlot} \textit{Top} shows, it is still rigidly expanding outwards in the rotation plane. Compared with the balanced 6D black rings  \cite{Kleihaus:2012xh, Dias:2014cia}, we find that the area and angular velocity are still 7\% below and 15\% above their respective equilibrium values in the final frame of our simulation. These values are consistent with the fact that the ring still has to expand by an additional 7\% in the rotation plane in order to reach the equilibrium $S^1$ radius while conserving angular momentum.

At late times, Fig. \ref{fig:thicknessPlot} shows that the flow in the $U<1$ region is unaffected by the pull from the outermost ring,
and the dynamics is therefore determined by the higher generation rings. These differ from the first generation in that they carry far too little angular momentum to be balanced. Instead, they are held in place by the tension of the membrane sections surrounding them. The tension of a membrane in 6D is proportional to its thickness, and different parts of the membrane have different thickness due to the pull from the outermost ring from the outset.
These differences are amplified as the GL instability develops on each of these local sections.
As the thicknesses of the surrounding membranes change, higher generation rings develop a radial velocity towards the
thicker sections.  This is clearly visible in Fig. \ref{fig:thicknessPlot} \textit{Middle}, and is large enough to significantly change the
width of a membrane section during the development of a new generation, thus affecting its formation time.

To obtain a precise value for the pinch-off time, we track the global minimum thickness of the membrane.
Even though the dynamics of higher generation rings prevents the formation of new generations from being self-similar,
the minimum thickness closely follows the scaling law
\begin{align} \label{eq:scalingLaw}
Z_\textrm{AH} = \alpha (t_c -t),
\end{align}
similar to black strings \cite{Lehner:2010pn,Lehner:2011wc} and the Rayleigh-Plateau instability of fluid columns
(see  Fig. \ref{fig:thicknessPlot} \textit{Bottom}).
This strongly supports our earlier conclusion that the BH will pinch off in finite asymptotic time, $t_c$, giving rise to a naked singularity. 
By performing a 2-parameter fit with \eqref{eq:scalingLaw}, we can obtain values for the pinch-off time $t_c$ and the
dimensionless constant $\alpha$. The value for the latter, $\alpha = (9.9 \pm 0.2)\times 10^{-3}$, is universal in that it is the same for all of our runs and is independent of the rotation parameter and initial data.

We may finally speculate about the end point. Fig. \ref{fig:thicknessPlot} suggests that after pinch-off the outermost ring will settle down to its balanced configuration, absorbing the nearby ($U \gtrsim 1$) membrane section.
Since the angular velocity in the central sections of the membrane is much too low to form balanced rings, sections closer to the centre will collapse into a spherical BH with negligible angular momentum. Therefore, the end point will be a black saturn in 6D. However, it will not be the saturn that maximizes the entropy for a given final mass and angular momentum, which consists of a central BH carrying all the mass surrounded by a thin ring that accounts for all the angular momentum \cite{Elvang:2007hg}. 
Nevertheless, we find that as $a/\mu^\frac{1}{3}$ is increased, the end point becomes more similar to this optimal configuration.

\PRLsection{Discussion}%
Our results provide evidence that the ultraspinning instability evolves into a naked singularity in finite asymptotic time,
and thus can be interpreted as a potential counterexample to the WCC in higher-dimensional asymptotically flat spaces. In
the approach to pinch-off, the minimum membrane thickness very closely follows a scaling
law \eqref{eq:scalingLaw}, with a universal constant $\alpha$. However, in $D\geq 6$ MP BHs are unstable to non-axisymmetric modes \cite{Shibata:2009ad,Shibata:2010wz,Dias:2014eua}. Therefore, to find a \textit{generic} violation of the WCC
in MP BH spacetimes, one has to consider the evolution under \textit{all} these instabilities. Work in this
direction is underway. Since the growth rate of the bar mode instability saturates \cite{Dias:2014eua}, we expect
that for sufficiently rapidly spinning MP BHs the ultraspinning instability will dominate. Once the first generation
ring has formed, the membrane inside becomes thinner by a factor of 50. Thus, the ultraspinning instability in this
paper is an order of magnitude faster than the axisymmetry-breaking GL instability of the outermost ring. Therefore, we
expect that for sufficiently large values of $a/\mu^\frac{1}{3}$, modes that preserve the axisymmetry are dominant in
all stages of the evolution, and that the violation of the WCC presented in this paper should be \textit{generic}.

\begin{acknowledgments}
We thank Roberto Emparan for discussions and Juha J\"aykk\"a and Kacper Kornet for their technical support.
We thank the \texttt{GRChombo} team (\url{http://grchombo.github.io/collaborators.html}) for the great collaboration.
Part of the computations for this Letter were undertaken on the COSMOS Shared Memory
system at DAMTP (University of Cambridge). COSMOS is operated on behalf of the STFC DiRAC HPC Facility and is funded by BIS National E-infrastructure capital Grant No. ST/ J005673/1 and STFC Grants No. ST/H008586/1, No. ST/ K00333X/1.
We thankfully acknowldege the computer resources at MareNostrum and the technical support provided by Barcelona
Supercomputing Center (FI-2016-3-0006 ``New frontiers in numerical general relativity"). P.F. and S.T. are supported by
the European Research Council Grant No. ERC-2014-StG 639022-NewNGR. P.F. is also supported by a Royal Society University
Research Fellowship (Grant No. UF140319). M.K. is
supported by an STFC studentship. L.L. is supported by NSERC, CIFAR and Perimeter Institute. This work has received
funding from the European Union's Horizon 2020 research and innovation programme under the Marie Sk\l odowska-Curie
Grant agreement No. 690904. P.F. and M.K. would like to thank Perimeter Institute for their hospitality during the final
stages of this work.  Research at Perimeter Institute is supported by the Government of Canada through the Department of Innovation, Science and Economic Development Canada, and by the Province of Ontario through the Ministry of Research and Innovation.
\phantom{\cite{Alcubierre:2002kk,AlcubierreBook,Smarr:1973zz}}
\end{acknowledgments}

\bibliography{refs}

\begin{thebibliography}{44}%
\makeatletter
\providecommand \@ifxundefined [1]{%
 \@ifx{#1\undefined}
}%
\providecommand \@ifnum [1]{%
 \ifnum #1\expandafter \@firstoftwo
 \else \expandafter \@secondoftwo
 \fi
}%
\providecommand \@ifx [1]{%
 \ifx #1\expandafter \@firstoftwo
 \else \expandafter \@secondoftwo
 \fi
}%
\providecommand \natexlab [1]{#1}%
\providecommand \enquote  [1]{``#1''}%
\providecommand \bibnamefont  [1]{#1}%
\providecommand \bibfnamefont [1]{#1}%
\providecommand \citenamefont [1]{#1}%
\providecommand \href@noop [0]{\@secondoftwo}%
\providecommand \href [0]{\begingroup \@sanitize@url \@href}%
\providecommand \@href[1]{\@@startlink{#1}\@@href}%
\providecommand \@@href[1]{\endgroup#1\@@endlink}%
\providecommand \@sanitize@url [0]{\catcode `\\12\catcode `\$12\catcode
  `\&12\catcode `\#12\catcode `\^12\catcode `\_12\catcode `\%12\relax}%
\providecommand \@@startlink[1]{}%
\providecommand \@@endlink[0]{}%
\providecommand \url  [0]{\begingroup\@sanitize@url \@url }%
\providecommand \@url [1]{\endgroup\@href {#1}{\urlprefix }}%
\providecommand \urlprefix  [0]{URL }%
\providecommand \Eprint [0]{\href }%
\providecommand \doibase [0]{http://dx.doi.org/}%
\providecommand \selectlanguage [0]{\@gobble}%
\providecommand \bibinfo  [0]{\@secondoftwo}%
\providecommand \bibfield  [0]{\@secondoftwo}%
\providecommand \translation [1]{[#1]}%
\providecommand \BibitemOpen [0]{}%
\providecommand \bibitemStop [0]{}%
\providecommand \bibitemNoStop [0]{.\EOS\space}%
\providecommand \EOS [0]{\spacefactor3000\relax}%
\providecommand \BibitemShut  [1]{\csname bibitem#1\endcsname}%
\let\auto@bib@innerbib\@empty
\bibitem [{\citenamefont {Abbott}\ \emph
  {et~al.}(2016{\natexlab{a}})\citenamefont {Abbott} \emph
  {et~al.}}]{Abbott:2016blz}%
  \BibitemOpen
  \bibfield  {author} {\bibinfo {author} {\bibfnamefont {B.~P.}\ \bibnamefont
  {Abbott}} \emph {et~al.},\ }\href {\doibase 10.1103/PhysRevLett.116.061102}
  {\bibfield  {journal} {\bibinfo  {journal} {Phys. Rev. Lett.}\ }\textbf
  {\bibinfo {volume} {116}},\ \bibinfo {pages} {061102} (\bibinfo {year}
  {2016}{\natexlab{a}})}\BibitemShut {NoStop}%
\bibitem [{\citenamefont {Abbott}\ \emph
  {et~al.}(2016{\natexlab{b}})\citenamefont {Abbott} \emph
  {et~al.}}]{Abbott:2016nmj}%
  \BibitemOpen
  \bibfield  {author} {\bibinfo {author} {\bibfnamefont {B.~P.}\ \bibnamefont
  {Abbott}} \emph {et~al.},\ }\href {\doibase 10.1103/PhysRevLett.116.241103}
  {\bibfield  {journal} {\bibinfo  {journal} {Phys. Rev. Lett.}\ }\textbf
  {\bibinfo {volume} {116}},\ \bibinfo {pages} {241103} (\bibinfo {year}
  {2016}{\natexlab{b}})}\BibitemShut {NoStop}%
\bibitem [{\citenamefont {Kerr}(1963)}]{Kerr:1963ud}%
  \BibitemOpen
  \bibfield  {author} {\bibinfo {author} {\bibfnamefont {R.~P.}\ \bibnamefont
  {Kerr}},\ }\href {\doibase 10.1103/PhysRevLett.11.237} {\bibfield  {journal}
  {\bibinfo  {journal} {Phys. Rev. Lett.}\ }\textbf {\bibinfo {volume} {11}},\
  \bibinfo {pages} {237} (\bibinfo {year} {1963})}\BibitemShut {NoStop}%
\bibitem [{\citenamefont {Aretakis}(2013)}]{Aretakis:2013dpa}%
  \BibitemOpen
  \bibfield  {author} {\bibinfo {author} {\bibfnamefont {S.}~\bibnamefont
  {Aretakis}},\ }\href {\doibase 10.1103/PhysRevD.87.084052} {\bibfield
  {journal} {\bibinfo  {journal} {Phys. Rev.}\ }\textbf {\bibinfo {volume}
  {D87}},\ \bibinfo {pages} {084052} (\bibinfo {year} {2013})},\ \Eprint
  {http://arxiv.org/abs/1304.4616} {arXiv:1304.4616 [gr-qc]} \BibitemShut
  {NoStop}%
\bibitem [{\citenamefont {Yang}\ \emph {et~al.}(2015)\citenamefont {Yang},
  \citenamefont {Zimmerman},\ and\ \citenamefont {Lehner}}]{Yang:2014tla}%
  \BibitemOpen
  \bibfield  {author} {\bibinfo {author} {\bibfnamefont {H.}~\bibnamefont
  {Yang}}, \bibinfo {author} {\bibfnamefont {A.}~\bibnamefont {Zimmerman}}, \
  and\ \bibinfo {author} {\bibfnamefont {L.}~\bibnamefont {Lehner}},\ }\href
  {\doibase 10.1103/PhysRevLett.114.081101} {\bibfield  {journal} {\bibinfo
  {journal} {Phys. Rev. Lett.}\ }\textbf {\bibinfo {volume} {114}},\ \bibinfo
  {pages} {081101} (\bibinfo {year} {2015})}\BibitemShut {NoStop}%
\bibitem [{\citenamefont {Gralla}\ \emph {et~al.}(2016)\citenamefont {Gralla},
  \citenamefont {Zimmerman},\ and\ \citenamefont {Zimmerman}}]{Gralla:2016sxp}%
  \BibitemOpen
  \bibfield  {author} {\bibinfo {author} {\bibfnamefont {S.~E.}\ \bibnamefont
  {Gralla}}, \bibinfo {author} {\bibfnamefont {A.}~\bibnamefont {Zimmerman}}, \
  and\ \bibinfo {author} {\bibfnamefont {P.}~\bibnamefont {Zimmerman}},\ }\href
  {\doibase 10.1103/PhysRevD.94.084017} {\bibfield  {journal} {\bibinfo
  {journal} {Phys. Rev.}\ }\textbf {\bibinfo {volume} {D94}},\ \bibinfo {pages}
  {084017} (\bibinfo {year} {2016})},\ \Eprint
  {http://arxiv.org/abs/1608.04739} {arXiv:1608.04739 [gr-qc]} \BibitemShut
  {NoStop}%
\bibitem [{\citenamefont {Gregory}\ and\ \citenamefont
  {Laflamme}(1993)}]{Gregory:1993vy}%
  \BibitemOpen
  \bibfield  {author} {\bibinfo {author} {\bibfnamefont {R.}~\bibnamefont
  {Gregory}}\ and\ \bibinfo {author} {\bibfnamefont {R.}~\bibnamefont
  {Laflamme}},\ }\href {\doibase 10.1103/PhysRevLett.70.2837} {\bibfield
  {journal} {\bibinfo  {journal} {Phys. Rev. Lett.}\ }\textbf {\bibinfo
  {volume} {70}},\ \bibinfo {pages} {2837} (\bibinfo {year}
  {1993})}\BibitemShut {NoStop}%
\bibitem [{\citenamefont {Lehner}\ and\ \citenamefont
  {Pretorius}(2010)}]{Lehner:2010pn}%
  \BibitemOpen
  \bibfield  {author} {\bibinfo {author} {\bibfnamefont {L.}~\bibnamefont
  {Lehner}}\ and\ \bibinfo {author} {\bibfnamefont {F.}~\bibnamefont
  {Pretorius}},\ }\href {\doibase 10.1103/PhysRevLett.105.101102} {\bibfield
  {journal} {\bibinfo  {journal} {Phys. Rev. Lett.}\ }\textbf {\bibinfo
  {volume} {105}},\ \bibinfo {pages} {101102} (\bibinfo {year}
  {2010})}\BibitemShut {NoStop}%
\bibitem [{\citenamefont {Emparan}\ and\ \citenamefont
  {Myers}(2003)}]{Emparan:2003sy}%
  \BibitemOpen
  \bibfield  {author} {\bibinfo {author} {\bibfnamefont {R.}~\bibnamefont
  {Emparan}}\ and\ \bibinfo {author} {\bibfnamefont {R.~C.}\ \bibnamefont
  {Myers}},\ }\href {\doibase 10.1088/1126-6708/2003/09/025} {\bibfield
  {journal} {\bibinfo  {journal} {JHEP}\ }\textbf {\bibinfo {volume} {09}},\
  \bibinfo {pages} {025} (\bibinfo {year} {2003})}\BibitemShut {NoStop}%
\bibitem [{\citenamefont {Emparan}\ and\ \citenamefont
  {Reall}(2002)}]{Emparan:2001wn}%
  \BibitemOpen
  \bibfield  {author} {\bibinfo {author} {\bibfnamefont {R.}~\bibnamefont
  {Emparan}}\ and\ \bibinfo {author} {\bibfnamefont {H.~S.}\ \bibnamefont
  {Reall}},\ }\href {\doibase 10.1103/PhysRevLett.88.101101} {\bibfield
  {journal} {\bibinfo  {journal} {Phys. Rev. Lett.}\ }\textbf {\bibinfo
  {volume} {88}},\ \bibinfo {pages} {101101} (\bibinfo {year}
  {2002})}\BibitemShut {NoStop}%
\bibitem [{\citenamefont {Elvang}\ \emph {et~al.}(2006)\citenamefont {Elvang},
  \citenamefont {Emparan},\ and\ \citenamefont {Virmani}}]{Elvang:2006dd}%
  \BibitemOpen
  \bibfield  {author} {\bibinfo {author} {\bibfnamefont {H.}~\bibnamefont
  {Elvang}}, \bibinfo {author} {\bibfnamefont {R.}~\bibnamefont {Emparan}}, \
  and\ \bibinfo {author} {\bibfnamefont {A.}~\bibnamefont {Virmani}},\ }\href
  {\doibase 10.1088/1126-6708/2006/12/074} {\bibfield  {journal} {\bibinfo
  {journal} {JHEP}\ }\textbf {\bibinfo {volume} {12}},\ \bibinfo {pages} {074}
  (\bibinfo {year} {2006})}\BibitemShut {NoStop}%
\bibitem [{\citenamefont {Figueras}\ \emph {et~al.}(2011)\citenamefont
  {Figueras}, \citenamefont {Murata},\ and\ \citenamefont
  {Reall}}]{Figueras:2011he}%
  \BibitemOpen
  \bibfield  {author} {\bibinfo {author} {\bibfnamefont {P.}~\bibnamefont
  {Figueras}}, \bibinfo {author} {\bibfnamefont {K.}~\bibnamefont {Murata}}, \
  and\ \bibinfo {author} {\bibfnamefont {H.~S.}\ \bibnamefont {Reall}},\ }\href
  {\doibase 10.1088/0264-9381/28/22/225030} {\bibfield  {journal} {\bibinfo
  {journal} {Class. Quant. Grav.}\ }\textbf {\bibinfo {volume} {28}},\ \bibinfo
  {pages} {225030} (\bibinfo {year} {2011})}\BibitemShut {NoStop}%
\bibitem [{\citenamefont {Santos}\ and\ \citenamefont
  {Way}(2015)}]{Santos:2015iua}%
  \BibitemOpen
  \bibfield  {author} {\bibinfo {author} {\bibfnamefont {J.~E.}\ \bibnamefont
  {Santos}}\ and\ \bibinfo {author} {\bibfnamefont {B.}~\bibnamefont {Way}},\
  }\href {\doibase 10.1103/PhysRevLett.114.221101} {\bibfield  {journal}
  {\bibinfo  {journal} {Phys. Rev. Lett.}\ }\textbf {\bibinfo {volume} {114}},\
  \bibinfo {pages} {221101} (\bibinfo {year} {2015})}\BibitemShut {NoStop}%
\bibitem [{\citenamefont {Tanabe}(2016)}]{Tanabe:2015hda}%
  \BibitemOpen
  \bibfield  {author} {\bibinfo {author} {\bibfnamefont {K.}~\bibnamefont
  {Tanabe}},\ }\href {\doibase 10.1007/JHEP02(2016)151} {\bibfield  {journal}
  {\bibinfo  {journal} {JHEP}\ }\textbf {\bibinfo {volume} {02}},\ \bibinfo
  {pages} {151} (\bibinfo {year} {2016})}\BibitemShut {NoStop}%
\bibitem [{\citenamefont {Figueras}\ \emph {et~al.}(2016)\citenamefont
  {Figueras}, \citenamefont {Kunesch},\ and\ \citenamefont
  {Tunyasuvunakool}}]{Figueras:2015hkb}%
  \BibitemOpen
  \bibfield  {author} {\bibinfo {author} {\bibfnamefont {P.}~\bibnamefont
  {Figueras}}, \bibinfo {author} {\bibfnamefont {M.}~\bibnamefont {Kunesch}}, \
  and\ \bibinfo {author} {\bibfnamefont {S.}~\bibnamefont {Tunyasuvunakool}},\
  }\href {\doibase 10.1103/PhysRevLett.116.071102} {\bibfield  {journal}
  {\bibinfo  {journal} {Phys. Rev. Lett.}\ }\textbf {\bibinfo {volume} {116}},\
  \bibinfo {pages} {071102} (\bibinfo {year} {2016})}\BibitemShut {NoStop}%
\bibitem [{\citenamefont {Tanabe}()}]{Tanabe:2016pjr}%
  \BibitemOpen
  \bibfield  {author} {\bibinfo {author} {\bibfnamefont {K.}~\bibnamefont
  {Tanabe}},\ }\href@noop {} {\ }\Eprint {http://arxiv.org/abs/1605.08116}
  {arXiv:1605.08116} \BibitemShut {NoStop}%
\bibitem [{\citenamefont {Dias}\ \emph {et~al.}(2009)\citenamefont {Dias},
  \citenamefont {Figueras}, \citenamefont {Monteiro}, \citenamefont {Santos},\
  and\ \citenamefont {Emparan}}]{Dias:2009iu}%
  \BibitemOpen
  \bibfield  {author} {\bibinfo {author} {\bibfnamefont {O.~J.~C.}\
  \bibnamefont {Dias}}, \bibinfo {author} {\bibfnamefont {P.}~\bibnamefont
  {Figueras}}, \bibinfo {author} {\bibfnamefont {R.}~\bibnamefont {Monteiro}},
  \bibinfo {author} {\bibfnamefont {J.~E.}\ \bibnamefont {Santos}}, \ and\
  \bibinfo {author} {\bibfnamefont {R.}~\bibnamefont {Emparan}},\ }\href
  {\doibase 10.1103/PhysRevD.80.111701} {\bibfield  {journal} {\bibinfo
  {journal} {Phys. Rev.}\ }\textbf {\bibinfo {volume} {D80}},\ \bibinfo {pages}
  {111701} (\bibinfo {year} {2009})}\BibitemShut {NoStop}%
\bibitem [{\citenamefont {Myers}\ and\ \citenamefont
  {Perry}(1986)}]{Myers:1986un}%
  \BibitemOpen
  \bibfield  {author} {\bibinfo {author} {\bibfnamefont {R.~C.}\ \bibnamefont
  {Myers}}\ and\ \bibinfo {author} {\bibfnamefont {M.~J.}\ \bibnamefont
  {Perry}},\ }\href@noop {} {\bibfield  {journal} {\bibinfo  {journal} {Ann.
  Phys.}\ }\textbf {\bibinfo {volume} {172}},\ \bibinfo {pages} {304} (\bibinfo
  {year} {1986})}\BibitemShut {NoStop}%
\bibitem [{Note1()}]{Note1}%
  \BibitemOpen
  \bibinfo {note} {MP BHs are the higher-dimensional analogues of Kerr
  BHs.}\BibitemShut {Stop}%
\bibitem [{\citenamefont {Dias}\ \emph
  {et~al.}(2014{\natexlab{a}})\citenamefont {Dias}, \citenamefont {Santos},\
  and\ \citenamefont {Way}}]{Dias:2014cia}%
  \BibitemOpen
  \bibfield  {author} {\bibinfo {author} {\bibfnamefont {O.~J.~C.}\
  \bibnamefont {Dias}}, \bibinfo {author} {\bibfnamefont {J.~E.}\ \bibnamefont
  {Santos}}, \ and\ \bibinfo {author} {\bibfnamefont {B.}~\bibnamefont {Way}},\
  }\href {\doibase 10.1007/JHEP07(2014)045} {\bibfield  {journal} {\bibinfo
  {journal} {JHEP}\ }\textbf {\bibinfo {volume} {07}},\ \bibinfo {pages} {045}
  (\bibinfo {year} {2014}{\natexlab{a}})}\BibitemShut {NoStop}%
\bibitem [{\citenamefont {Emparan}\ \emph {et~al.}(2014)\citenamefont
  {Emparan}, \citenamefont {Figueras},\ and\ \citenamefont
  {Martinez}}]{Emparan:2014pra}%
  \BibitemOpen
  \bibfield  {author} {\bibinfo {author} {\bibfnamefont {R.}~\bibnamefont
  {Emparan}}, \bibinfo {author} {\bibfnamefont {P.}~\bibnamefont {Figueras}}, \
  and\ \bibinfo {author} {\bibfnamefont {M.}~\bibnamefont {Martinez}},\ }\href
  {\doibase 10.1007/JHEP12(2014)072} {\bibfield  {journal} {\bibinfo  {journal}
  {JHEP}\ }\textbf {\bibinfo {volume} {12}},\ \bibinfo {pages} {072} (\bibinfo
  {year} {2014})}\BibitemShut {NoStop}%
\bibitem [{\citenamefont {Pretorius}(2005)}]{Pretorius:2004jg}%
  \BibitemOpen
  \bibfield  {author} {\bibinfo {author} {\bibfnamefont {F.}~\bibnamefont
  {Pretorius}},\ }\href {\doibase 10.1088/0264-9381/22/2/014} {\bibfield
  {journal} {\bibinfo  {journal} {Class. Quant. Grav.}\ }\textbf {\bibinfo
  {volume} {22}},\ \bibinfo {pages} {425} (\bibinfo {year} {2005})}\BibitemShut
  {NoStop}%
\bibitem [{\citenamefont {Shibata}\ and\ \citenamefont
  {Yoshino}(2010{\natexlab{a}})}]{Shibata:2010wz}%
  \BibitemOpen
  \bibfield  {author} {\bibinfo {author} {\bibfnamefont {M.}~\bibnamefont
  {Shibata}}\ and\ \bibinfo {author} {\bibfnamefont {H.}~\bibnamefont
  {Yoshino}},\ }\href {\doibase 10.1103/PhysRevD.81.104035} {\bibfield
  {journal} {\bibinfo  {journal} {Phys. Rev.}\ }\textbf {\bibinfo {volume}
  {D81}},\ \bibinfo {pages} {104035} (\bibinfo {year}
  {2010}{\natexlab{a}})}\BibitemShut {NoStop}%
\bibitem [{\citenamefont {Cook}\ \emph {et~al.}(2016)\citenamefont {Cook},
  \citenamefont {Figueras}, \citenamefont {Kunesch}, \citenamefont {Sperhake},\
  and\ \citenamefont {Tunyasuvunakool}}]{Cook:2016soy}%
  \BibitemOpen
  \bibfield  {author} {\bibinfo {author} {\bibfnamefont {W.~G.}\ \bibnamefont
  {Cook}}, \bibinfo {author} {\bibfnamefont {P.}~\bibnamefont {Figueras}},
  \bibinfo {author} {\bibfnamefont {M.}~\bibnamefont {Kunesch}}, \bibinfo
  {author} {\bibfnamefont {U.}~\bibnamefont {Sperhake}}, \ and\ \bibinfo
  {author} {\bibfnamefont {S.}~\bibnamefont {Tunyasuvunakool}},\ }\href
  {\doibase 10.1142/S0218271816410133} {\bibfield  {journal} {\bibinfo
  {journal} {Int. J. Mod. Phys.}\ }\textbf {\bibinfo {volume} {D25}},\ \bibinfo
  {pages} {1641013} (\bibinfo {year} {2016})}\BibitemShut {NoStop}%
\bibitem [{\citenamefont {Alic}\ \emph {et~al.}(2012)\citenamefont {Alic},
  \citenamefont {Bona-Casas}, \citenamefont {Bona}, \citenamefont {Rezzolla},\
  and\ \citenamefont {Palenzuela}}]{Alic:2011gg}%
  \BibitemOpen
  \bibfield  {author} {\bibinfo {author} {\bibfnamefont {D.}~\bibnamefont
  {Alic}}, \bibinfo {author} {\bibfnamefont {C.}~\bibnamefont {Bona-Casas}},
  \bibinfo {author} {\bibfnamefont {C.}~\bibnamefont {Bona}}, \bibinfo {author}
  {\bibfnamefont {L.}~\bibnamefont {Rezzolla}}, \ and\ \bibinfo {author}
  {\bibfnamefont {C.}~\bibnamefont {Palenzuela}},\ }\href {\doibase
  10.1103/PhysRevD.85.064040} {\bibfield  {journal} {\bibinfo  {journal} {Phys.
  Rev.}\ }\textbf {\bibinfo {volume} {D85}},\ \bibinfo {pages} {064040}
  (\bibinfo {year} {2012})}\BibitemShut {NoStop}%
\bibitem [{\citenamefont {Weyhausen}\ \emph {et~al.}(2012)\citenamefont
  {Weyhausen}, \citenamefont {Bernuzzi},\ and\ \citenamefont
  {Hilditch}}]{Weyhausen:2011cg}%
  \BibitemOpen
  \bibfield  {author} {\bibinfo {author} {\bibfnamefont {A.}~\bibnamefont
  {Weyhausen}}, \bibinfo {author} {\bibfnamefont {S.}~\bibnamefont {Bernuzzi}},
  \ and\ \bibinfo {author} {\bibfnamefont {D.}~\bibnamefont {Hilditch}},\
  }\href {\doibase 10.1103/PhysRevD.85.024038} {\bibfield  {journal} {\bibinfo
  {journal} {Phys. Rev.}\ }\textbf {\bibinfo {volume} {D85}},\ \bibinfo {pages}
  {024038} (\bibinfo {year} {2012})}\BibitemShut {NoStop}%
\bibitem [{\citenamefont {Alic}\ \emph {et~al.}(2013)\citenamefont {Alic},
  \citenamefont {Kastaun},\ and\ \citenamefont {Rezzolla}}]{Alic:2013xsa}%
  \BibitemOpen
  \bibfield  {author} {\bibinfo {author} {\bibfnamefont {D.}~\bibnamefont
  {Alic}}, \bibinfo {author} {\bibfnamefont {W.}~\bibnamefont {Kastaun}}, \
  and\ \bibinfo {author} {\bibfnamefont {L.}~\bibnamefont {Rezzolla}},\ }\href
  {\doibase 10.1103/PhysRevD.88.064049} {\bibfield  {journal} {\bibinfo
  {journal} {Phys. Rev.}\ }\textbf {\bibinfo {volume} {D88}},\ \bibinfo {pages}
  {064049} (\bibinfo {year} {2013})}\BibitemShut {NoStop}%
\bibitem [{Sup()}]{Supplemental}%
  \BibitemOpen
  \href@noop {} {}\bibinfo {note} {See Supplemental Material for a description
  of our modified Gamma-driver, technical details of our apparent horizon
  finder, and convergence test results, which includes Refs.
  [42-44].}\BibitemShut {Stop}%
\bibitem [{\citenamefont {Brown}\ \emph {et~al.}(2007)\citenamefont {Brown},
  \citenamefont {Sarbach}, \citenamefont {Schnetter}, \citenamefont {Tiglio},
  \citenamefont {Diener}, \citenamefont {Hawke},\ and\ \citenamefont
  {Pollney}}]{Brown:2007pg}%
  \BibitemOpen
  \bibfield  {author} {\bibinfo {author} {\bibfnamefont {J.~D.}\ \bibnamefont
  {Brown}}, \bibinfo {author} {\bibfnamefont {O.}~\bibnamefont {Sarbach}},
  \bibinfo {author} {\bibfnamefont {E.}~\bibnamefont {Schnetter}}, \bibinfo
  {author} {\bibfnamefont {M.}~\bibnamefont {Tiglio}}, \bibinfo {author}
  {\bibfnamefont {P.}~\bibnamefont {Diener}}, \bibinfo {author} {\bibfnamefont
  {I.}~\bibnamefont {Hawke}}, \ and\ \bibinfo {author} {\bibfnamefont
  {D.}~\bibnamefont {Pollney}},\ }\href {\doibase 10.1103/PhysRevD.76.081503}
  {\bibfield  {journal} {\bibinfo  {journal} {Phys. Rev.}\ }\textbf {\bibinfo
  {volume} {D76}},\ \bibinfo {pages} {081503} (\bibinfo {year}
  {2007})}\BibitemShut {NoStop}%
\bibitem [{\citenamefont {Brown}\ \emph {et~al.}(2009)\citenamefont {Brown},
  \citenamefont {Diener}, \citenamefont {Sarbach}, \citenamefont {Schnetter},\
  and\ \citenamefont {Tiglio}}]{Brown:2008sb}%
  \BibitemOpen
  \bibfield  {author} {\bibinfo {author} {\bibfnamefont {J.~D.}\ \bibnamefont
  {Brown}}, \bibinfo {author} {\bibfnamefont {P.}~\bibnamefont {Diener}},
  \bibinfo {author} {\bibfnamefont {O.}~\bibnamefont {Sarbach}}, \bibinfo
  {author} {\bibfnamefont {E.}~\bibnamefont {Schnetter}}, \ and\ \bibinfo
  {author} {\bibfnamefont {M.}~\bibnamefont {Tiglio}},\ }\href {\doibase
  10.1103/PhysRevD.79.044023} {\bibfield  {journal} {\bibinfo  {journal} {Phys.
  Rev.}\ }\textbf {\bibinfo {volume} {D79}},\ \bibinfo {pages} {044023}
  (\bibinfo {year} {2009})}\BibitemShut {NoStop}%
\bibitem [{\citenamefont {Thornburg}(2007)}]{Thornburg:2006zb}%
  \BibitemOpen
  \bibfield  {author} {\bibinfo {author} {\bibfnamefont {J.}~\bibnamefont
  {Thornburg}},\ }\href@noop {} {\bibfield  {journal} {\bibinfo  {journal}
  {Living Rev. Rel.}\ }\textbf {\bibinfo {volume} {10}},\ \bibinfo {pages} {3}
  (\bibinfo {year} {2007})},\ \Eprint {http://arxiv.org/abs/gr-qc/0512169}
  {arXiv:gr-qc/0512169 [gr-qc]} \BibitemShut {NoStop}%
\bibitem [{\citenamefont {Tunyasuvunakool}(2016)}]{SaranThesis}%
  \BibitemOpen
  \bibfield  {author} {\bibinfo {author} {\bibfnamefont {S.}~\bibnamefont
  {Tunyasuvunakool}},\ }\href {\doibase 10.17863/CAM.7743} {\bibfield
  {journal} {\bibinfo  {journal} {PhD Thesis, University of Cambridge}\ }
  (\bibinfo {year} {2016}),\ 10.17863/CAM.7743}\BibitemShut {NoStop}%
\bibitem [{\citenamefont {Clough}\ \emph {et~al.}(2015)\citenamefont {Clough},
  \citenamefont {Figueras}, \citenamefont {Finkel}, \citenamefont {Kunesch},
  \citenamefont {Lim},\ and\ \citenamefont {Tunyasuvunakool}}]{Clough:2015sqa}%
  \BibitemOpen
  \bibfield  {author} {\bibinfo {author} {\bibfnamefont {K.}~\bibnamefont
  {Clough}}, \bibinfo {author} {\bibfnamefont {P.}~\bibnamefont {Figueras}},
  \bibinfo {author} {\bibfnamefont {H.}~\bibnamefont {Finkel}}, \bibinfo
  {author} {\bibfnamefont {M.}~\bibnamefont {Kunesch}}, \bibinfo {author}
  {\bibfnamefont {E.~A.}\ \bibnamefont {Lim}}, \ and\ \bibinfo {author}
  {\bibfnamefont {S.}~\bibnamefont {Tunyasuvunakool}},\ }\href {\doibase
  10.1088/0264-9381/32/24/245011} {\bibfield  {journal} {\bibinfo  {journal}
  {Class. Quant. Grav.}\ }\textbf {\bibinfo {volume} {32}},\ \bibinfo {pages}
  {245011} (\bibinfo {year} {2015})}\BibitemShut {NoStop}%
\bibitem [{\citenamefont {Adams}\ \emph {et~al.}(2015)\citenamefont {Adams},
  \citenamefont {Colella}, \citenamefont {Graves}, \citenamefont {Johnson},
  \citenamefont {Keen}, \citenamefont {Ligocki}, \citenamefont {Martin},
  \citenamefont {McCorquodale}, \citenamefont {Modiano}, \citenamefont
  {Schwartz}, \citenamefont {Sternberg},\ and\ \citenamefont
  {Van~Straalen}}]{chombo-design-doc}%
  \BibitemOpen
  \bibfield  {author} {\bibinfo {author} {\bibfnamefont {M.}~\bibnamefont
  {Adams}}, \bibinfo {author} {\bibfnamefont {P.}~\bibnamefont {Colella}},
  \bibinfo {author} {\bibfnamefont {D.}~\bibnamefont {Graves}}, \bibinfo
  {author} {\bibfnamefont {J.}~\bibnamefont {Johnson}}, \bibinfo {author}
  {\bibfnamefont {N.}~\bibnamefont {Keen}}, \bibinfo {author} {\bibfnamefont
  {T.}~\bibnamefont {Ligocki}}, \bibinfo {author} {\bibfnamefont
  {D.}~\bibnamefont {Martin}}, \bibinfo {author} {\bibfnamefont
  {P.}~\bibnamefont {McCorquodale}}, \bibinfo {author} {\bibfnamefont
  {D.}~\bibnamefont {Modiano}}, \bibinfo {author} {\bibfnamefont
  {P.}~\bibnamefont {Schwartz}}, \bibinfo {author} {\bibfnamefont
  {T.}~\bibnamefont {Sternberg}}, \ and\ \bibinfo {author} {\bibfnamefont
  {B.}~\bibnamefont {Van~Straalen}},\ }\href
  {http://crd.lbl.gov/assets/pubs_presos/chomboDesign.pdf} {\emph {\bibinfo
  {title} {Chombo Software Package for AMR Applications - Design Document}}},\
  \bibinfo {type} {Tech. Rep.}\ \bibinfo {number} {LBNL-6616E}\ (\bibinfo
  {institution} {Lawrence Berkeley National Laboratory},\ \bibinfo {year}
  {2015})\BibitemShut {NoStop}%
\bibitem [{Note2()}]{Note2}%
  \BibitemOpen
  \bibinfo {note} {Videos can be found at \protect \url
  {http://grchombo.github.io}.}\BibitemShut {Stop}%
\bibitem [{\citenamefont {Hovdebo}\ and\ \citenamefont
  {Myers}(2006)}]{Hovdebo:2006jy}%
  \BibitemOpen
  \bibfield  {author} {\bibinfo {author} {\bibfnamefont {J.~L.}\ \bibnamefont
  {Hovdebo}}\ and\ \bibinfo {author} {\bibfnamefont {R.~C.}\ \bibnamefont
  {Myers}},\ }\href@noop {} {\bibfield  {journal} {\bibinfo  {journal} {Phys.
  Rev.}\ }\textbf {\bibinfo {volume} {D73}},\ \bibinfo {pages} {084013}
  (\bibinfo {year} {2006})}\BibitemShut {NoStop}%
\bibitem [{\citenamefont {Kleihaus}\ \emph {et~al.}(2013)\citenamefont
  {Kleihaus}, \citenamefont {Kunz},\ and\ \citenamefont
  {Radu}}]{Kleihaus:2012xh}%
  \BibitemOpen
  \bibfield  {author} {\bibinfo {author} {\bibfnamefont {B.}~\bibnamefont
  {Kleihaus}}, \bibinfo {author} {\bibfnamefont {J.}~\bibnamefont {Kunz}}, \
  and\ \bibinfo {author} {\bibfnamefont {E.}~\bibnamefont {Radu}},\ }\href
  {\doibase 10.1016/j.physletb.2012.11.015} {\bibfield  {journal} {\bibinfo
  {journal} {Phys. Lett.}\ }\textbf {\bibinfo {volume} {B718}},\ \bibinfo
  {pages} {1073} (\bibinfo {year} {2013})},\ \Eprint
  {http://arxiv.org/abs/1205.5437} {arXiv:1205.5437 [hep-th]} \BibitemShut
  {NoStop}%
\bibitem [{\citenamefont {Lehner}\ and\ \citenamefont
  {Pretorius}(2011)}]{Lehner:2011wc}%
  \BibitemOpen
  \bibfield  {author} {\bibinfo {author} {\bibfnamefont {L.}~\bibnamefont
  {Lehner}}\ and\ \bibinfo {author} {\bibfnamefont {F.}~\bibnamefont
  {Pretorius}},\ }\href@noop {} {\  (\bibinfo {year} {2011})},\ \Eprint
  {http://arxiv.org/abs/1106.5184} {arXiv:1106.5184 [gr-qc]} \BibitemShut
  {NoStop}%
\bibitem [{\citenamefont {Elvang}\ \emph {et~al.}(2007)\citenamefont {Elvang},
  \citenamefont {Emparan},\ and\ \citenamefont {Figueras}}]{Elvang:2007hg}%
  \BibitemOpen
  \bibfield  {author} {\bibinfo {author} {\bibfnamefont {H.}~\bibnamefont
  {Elvang}}, \bibinfo {author} {\bibfnamefont {R.}~\bibnamefont {Emparan}}, \
  and\ \bibinfo {author} {\bibfnamefont {P.}~\bibnamefont {Figueras}},\
  }\href@noop {} {\bibfield  {journal} {\bibinfo  {journal} {JHEP}\ }\textbf
  {\bibinfo {volume} {05}},\ \bibinfo {pages} {056} (\bibinfo {year}
  {2007})}\BibitemShut {NoStop}%
\bibitem [{\citenamefont {Shibata}\ and\ \citenamefont
  {Yoshino}(2010{\natexlab{b}})}]{Shibata:2009ad}%
  \BibitemOpen
  \bibfield  {author} {\bibinfo {author} {\bibfnamefont {M.}~\bibnamefont
  {Shibata}}\ and\ \bibinfo {author} {\bibfnamefont {H.}~\bibnamefont
  {Yoshino}},\ }\href {\doibase 10.1103/PhysRevD.81.021501} {\bibfield
  {journal} {\bibinfo  {journal} {Phys. Rev.}\ }\textbf {\bibinfo {volume}
  {D81}},\ \bibinfo {pages} {021501} (\bibinfo {year}
  {2010}{\natexlab{b}})}\BibitemShut {NoStop}%
\bibitem [{\citenamefont {Dias}\ \emph
  {et~al.}(2014{\natexlab{b}})\citenamefont {Dias}, \citenamefont {Hartnett},\
  and\ \citenamefont {Santos}}]{Dias:2014eua}%
  \BibitemOpen
  \bibfield  {author} {\bibinfo {author} {\bibfnamefont {O.~J.~C.}\
  \bibnamefont {Dias}}, \bibinfo {author} {\bibfnamefont {G.~S.}\ \bibnamefont
  {Hartnett}}, \ and\ \bibinfo {author} {\bibfnamefont {J.~E.}\ \bibnamefont
  {Santos}},\ }\href {\doibase 10.1088/0264-9381/31/24/245011} {\bibfield
  {journal} {\bibinfo  {journal} {Class. Quant. Grav.}\ }\textbf {\bibinfo
  {volume} {31}},\ \bibinfo {pages} {245011} (\bibinfo {year}
  {2014}{\natexlab{b}})}\BibitemShut {NoStop}%
\bibitem [{\citenamefont {Alcubierre}\ \emph {et~al.}(2003)\citenamefont
  {Alcubierre}, \citenamefont {Bruegmann}, \citenamefont {Diener},
  \citenamefont {Koppitz}, \citenamefont {Pollney}, \citenamefont {Seidel},\
  and\ \citenamefont {Takahashi}}]{Alcubierre:2002kk}%
  \BibitemOpen
  \bibfield  {author} {\bibinfo {author} {\bibfnamefont {M.}~\bibnamefont
  {Alcubierre}}, \bibinfo {author} {\bibfnamefont {B.}~\bibnamefont
  {Bruegmann}}, \bibinfo {author} {\bibfnamefont {P.}~\bibnamefont {Diener}},
  \bibinfo {author} {\bibfnamefont {M.}~\bibnamefont {Koppitz}}, \bibinfo
  {author} {\bibfnamefont {D.}~\bibnamefont {Pollney}}, \bibinfo {author}
  {\bibfnamefont {E.}~\bibnamefont {Seidel}}, \ and\ \bibinfo {author}
  {\bibfnamefont {R.}~\bibnamefont {Takahashi}},\ }\href {\doibase
  10.1103/PhysRevD.67.084023} {\bibfield  {journal} {\bibinfo  {journal} {Phys.
  Rev.}\ }\textbf {\bibinfo {volume} {D67}},\ \bibinfo {pages} {084023}
  (\bibinfo {year} {2003})},\ \Eprint {http://arxiv.org/abs/gr-qc/0206072}
  {arXiv:gr-qc/0206072 [gr-qc]} \BibitemShut {NoStop}%
\bibitem [{\citenamefont {Alcubierre}(2008)}]{AlcubierreBook}%
  \BibitemOpen
  \bibfield  {author} {\bibinfo {author} {\bibfnamefont {M.}~\bibnamefont
  {Alcubierre}},\ }\href@noop {} {\bibfield  {journal} {\bibinfo  {journal}
  {Oxford University Press}\ } (\bibinfo {year} {2008})}\BibitemShut {NoStop}%
\bibitem [{\citenamefont {Smarr}(1973)}]{Smarr:1973zz}%
  \BibitemOpen
  \bibfield  {author} {\bibinfo {author} {\bibfnamefont {L.}~\bibnamefont
  {Smarr}},\ }\href {\doibase 10.1103/PhysRevD.7.289} {\bibfield  {journal}
  {\bibinfo  {journal} {Phys. Rev.}\ }\textbf {\bibinfo {volume} {D7}},\
  \bibinfo {pages} {289} (\bibinfo {year} {1973})}\BibitemShut {NoStop}%
\end{thebibliography}%
\bibliographystyle{apsrev4-1}

\pagebreak

\makeatletter 
\renewcommand{\thefigure}{S\@arabic\c@figure}
\makeatother

\makeatletter 
\renewcommand{\theequation}{S\@arabic\c@equation}
\makeatother

\onecolumngrid

\begin{figure*}
{\large\bf Supplemental Material}
\end{figure*}

\twocolumngrid

\setcounter{figure}{0}
\setcounter{equation}{0}

\PRLsection{Modified Gamma-Driver}
In this section, we provide more details on the modified Gamma-Driver shift condition employed in our simulations. We essentially followed \cite{Figueras:2015hkb}.

Since the initial data for rapidly spinning MP BHs is far from being conformally flat, most grid variables become very
large inside the black hole.
This means that we cannot use the standard Gamma-Driver shift condition \cite{Alcubierre:2002kk} to evolve the shift as it causes the large
initial values of the conformal connection functions, $\hat \Gamma^i$, to freeze in, even when advection
terms are included.
This behavior can be understood by considering the integrated form of the Gamma-Driver
\begin{align}
\partial_t \beta^i - F\left(\partial_t \beta^i\right)_{t=0} = F(\hat \Gamma^i - \hat \Gamma^i_{t=0}) - \eta (\beta^i -
\beta^i_{t=0}),
\end{align}
where $F$ and $\eta$ are free gauge parameters.
If $F$ is chosen such that the Gamma-Driver successfully counters the stretching of slices around the black hole, the
time-scale over which the shift settles down to an approximately steady state is much faster than the time-scale of the
evolution. If $\hat \Gamma^i_{t=0}$ is very large compared to $\left(\partial_t \beta^i\right)_{t=0}$ and $\beta^i_{t=0}$
inside the black hole, this
steady state requires that $\hat \Gamma^i$ remains frozen at its initial value.
One solution would be to cancel the offending integration constant $\hat \Gamma^i_{t=0}$ using the initial value for
$\beta^i$ or $\partial_t \beta^i$. However, this causes an unacceptably fast gauge adjustment at early times.

Instead, we evolve the shift using
\begin{align} \label{eq:GammaDriverModified}
\partial_t \beta^i = F(\hat  \Gamma^i - f(t) \hat  \Gamma^i_{t=0}) - \eta (\beta^i - \beta^i_{t=0}) + \beta^k
\partial_k \beta^i,
\end{align}
where $f(t)$ is a function that is identically equal to $1$ initially and then decays in time.
This gently unfreezes the initial value of $\hat  \Gamma^i$, allowing it to
tend to zero.
For our simulations we used $F=0.6$, $\eta = 1$ and
\begin{align} \label{eq:foft}
f(t) = \exp \left[ - \left(\delta_1 r_{\mathrm{hor}}^2/r^2 + \delta_2\right)
t^2/\mu^\frac{2}{3} \right]\,,
\end{align}
where $r_\mathrm{hor}$ is the location of the horizon, and
$\delta_1$ and $\delta_2$ are two dimensionless parameters that we chose to be $0.2$ and $0.075$ respectively.
The purpose of the first term in the exponential in \eqref{eq:foft}
is to speed up the gauge adjustment deep inside the apparent horizon, where the initial value of
$\hat \Gamma^i$ is larger and constraint violations do not matter.
Note that since \texttt{GRChombo} is cell-centered, \eqref{eq:foft} is never evaluated at $r=0$.

\PRLsection{Apparent Horizons}
In this section, we provide further details about our apparent horizon finder. A more thorough discussion can be found in \cite{SaranThesis}.

Consider a $d$-dimensional constant-time slice $\Sigma$ in the full spacetime. The apparent horizon $\mathcal H$ is
defined as the outermost marginally trapped surface on $\Sigma$. Even though $\mathcal H$ is gauge dependent, as it
depends on the particular choice of slicing, in equilibrium spacetimes it coincides with the event horizon. Therefore,
as the system approaches equilibrium, $\mathcal H$ should approach the event horizon. This is of special relevance in
the present setting: as the ultraspinning instability unfolds, the dynamics happens at ever decreasing length scales.
This implies that the geometry is in quasi-equilibrium  almost everywhere,
except for very small regions. Therefore, we expect that by the time the pinch off happens, $\mathcal H$ is very close to the event horizon.

Recall that on $\mathcal H$, the expansion of the outgoing null geodesics vanishes:
\begin{equation}
\Theta = (\gamma^{ab} - s^as^b)(-k_{ab} - K_{ab}) \equiv 0\,,
\label{eq:expansion}
\end{equation}
where $\gamma_{ab}$ and $K_{ab}$ are the induced metric and the extrinsic curvature of $\Sigma$, respectively. Here $s^a$ is the outward unit normal to $\mathcal H$ in $\Sigma$, $k_{ab} = -\frac{1}{2}\mathcal L_s h_{ab}$ is the extrinsic curvature of $\mathcal H$ in $\Sigma$, and $h_{ab}$ is the induced metric on $\mathcal H$. See \cite{AlcubierreBook} for a standard derivation.

The standard approach in numerical relativity is to parameterize $\mathcal H$ in terms of a scalar function $F$ such that $\mathcal H$ corresponds to the zero contour,
\begin{equation}
x^1 - F(x^2,\ldots,x^{d-1}) = 0\,,
\label{eq:levelset}
\end{equation}
where $x^i$ are coordinates on $\Sigma$. The main limitation of this approach is that \eqref{eq:levelset} can only
describe surfaces that occupy a convex region in the $(x^2,\ldots,x^{d-1})$--hyperplane.
Unless we make an unreasonably complicated choice of coordinates, $x^i$, this assumption breaks down in
the final stages of the evolution of the ultraspinning instability of MP black holes or in the GL instability of black
rings.

To describe more general apparent horizons, we treat $\mathcal H$ as a general parametric surface. More precisely, we define $\mathcal H$ via
\begin{equation}
x^i = X^i(u^\alpha)\,,
\end{equation}
where $u^\alpha$, $\alpha = 1,\ldots,d-1$, are parameters on $\mathcal H$. Our goal is to determine the $d$ unknown functions $X^i$ that specify $\mathcal H$. The tangent and normal vectors to $\mathcal H$ in $\Sigma$ are given by
\begin{equation}
T^i_{(\alpha)} = \frac{\partial X^i}{\partial u^\alpha}\,,
\quad S^i = \star\left(T_{(1)}\wedge \cdots \wedge T_{(d-1)}\right)_j \gamma^{ji}\,,
\end{equation}
respectively. Let $t^i_{(\alpha)}$ and $s^i$ denote the corresponding unit vectors. Then, the extrinsic curvature of $\mathcal H$ in $\Sigma$ is given by
\begin{equation}
\begin{aligned}
k_{\alpha\beta} =& - T^i_{(\alpha)}T^j_{(\beta)}\nabla_j s_i \\
=&~s_i\left(\frac{\partial^2 X^i}{\partial u^\alpha \partial u^\beta} + \Gamma^i_{\phantom i j k } \frac{\partial X^j}{\partial u^\alpha} \frac{\partial X^k}{\partial u^\beta}\right) \,.
\end{aligned}
\end{equation}
Denoting the induced metric on $\mathcal H$ by $h_{\alpha\beta} = T^i_{(\alpha)}T^j_{(\beta)}\gamma_{ij}$, equation \eqref{eq:expansion} becomes
\begin{equation}
s_i\,h^{\alpha\beta}\left(\frac{\partial^2 X^i}{\partial u^\alpha \partial u^\beta} + \Gamma^{i}_{\phantom i j k}\frac{\partial X^j}{\partial u^\alpha}\frac{\partial X^k}{\partial u^\beta}\right) + (\gamma^{ij} - s^i\,s^j)K_{ij} = 0\,,
\label{eq:expansion2}
\end{equation}
where we treat $s^i$ as a function of the $\partial X^i/\partial u^\alpha$.

Since we have $d$ unknowns, we must provide $d-1$ additional equations in order to complete the system of equations. The
latter simply correspond to the gauge fixing conditions for the parameters $u^\alpha$. One option to fix this gauge freedom
is to impose a generalized harmonic gauge condition,
\begin{equation}
\Box_{\mathcal H}u^\alpha = H^\alpha(u^\beta),
\label{eq:preGH}
\end{equation}
where $H^\alpha$ are some suitably chosen source functions. For the numerical implementation it turns out to be more convenient to implement \eqref{eq:preGH} with the index lowered (with the induced metric on $\mathcal H$). Thus, expanding \eqref{eq:preGH} with lowered indices, we arrive at our proposed gauge fixing conditions:
\begin{equation}
\gamma_{ij}\,h^{\beta \gamma}\left(\frac{\partial^2 X^i}{\partial u^\beta\partial u^\gamma} \frac{\partial X^j}{\partial u^\alpha} + \Gamma^i_{\phantom i j k}\frac{\partial X^j}{\partial u^\alpha}\frac{\partial X^k}{\partial u^\beta}\frac{\partial X^l}{\partial u^\gamma}\right) = H_{\alpha}(u^\beta)\,,
\label{eq:GH}
\end{equation}
where $H_{\alpha}$ are some $d-1$ prescribed functions. We can rewrite \eqref{eq:expansion2} and \eqref{eq:GH} in a form that makes their common structure manifest:
\begin{align}
&\gamma_{ij}\,h^{\beta \gamma}\left(\frac{\partial^2 X^i}{\partial u^\beta\partial u^\gamma} S^j + \Gamma^i_{\phantom i j k} S^j T^k_{(\beta)}T^l_{(\gamma)}\right) =  \parallel S^k \parallel (\gamma^{ij} - s^i s^j)K_{ij}\,,
\label{eq:final1} \\
&\gamma_{ij}\,h^{\beta \gamma}\left(\frac{\partial^2 X^i}{\partial u^\beta\partial u^\gamma} T^j_{(\alpha)} + \Gamma^i_{\phantom i j k} T^j_{(\alpha)} T^k_{(\beta)} T^l_{(\gamma)}\right) = H_\alpha\,.
\label{eq:final2}
\end{align}
Equations \eqref{eq:final1} and \eqref{eq:final2} are manifestly elliptic, and their solution determines $\mathcal H$ in a general situation.

\begin{figure}
\centering
\includegraphics[width=\columnwidth]{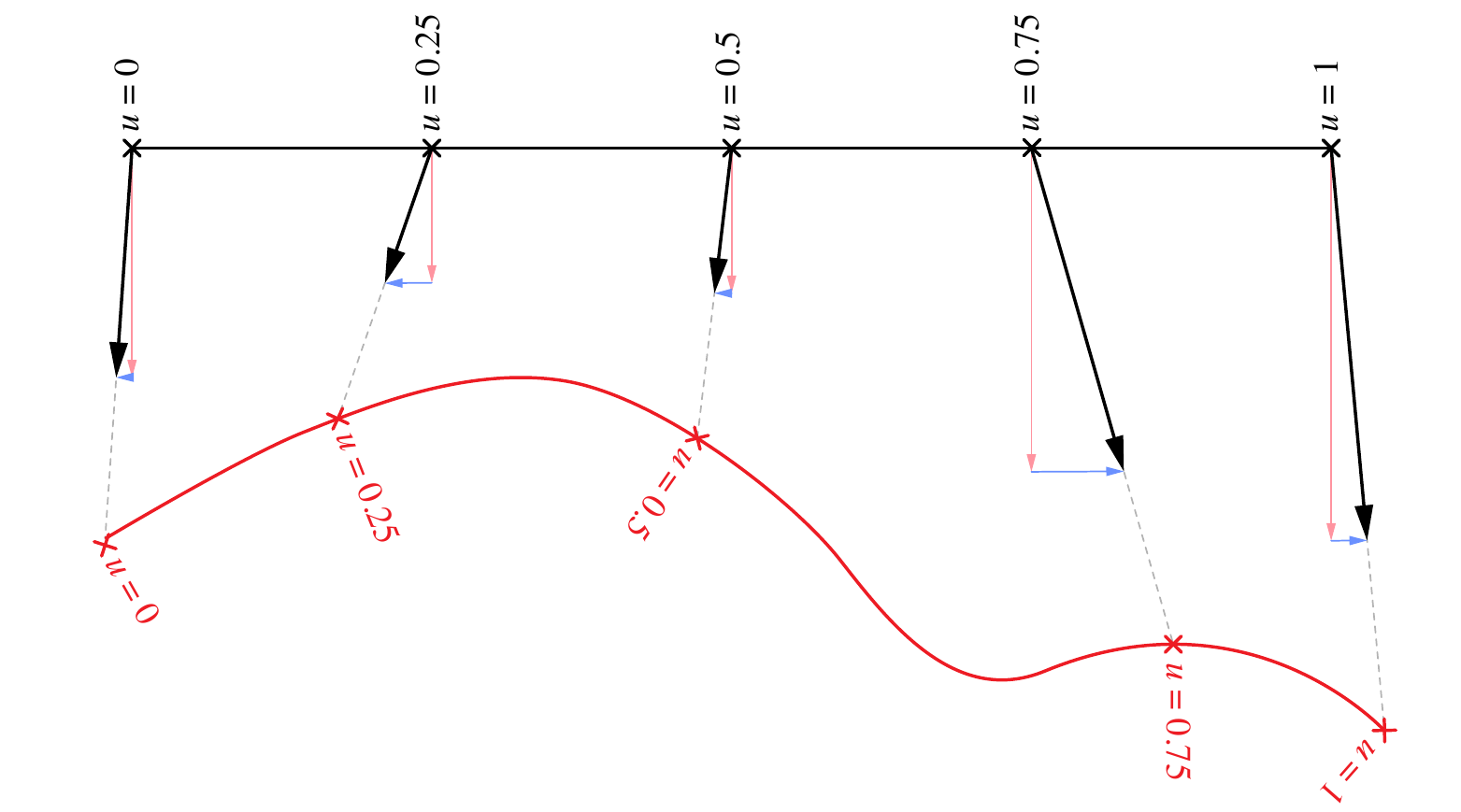}
\includegraphics[width=\columnwidth]{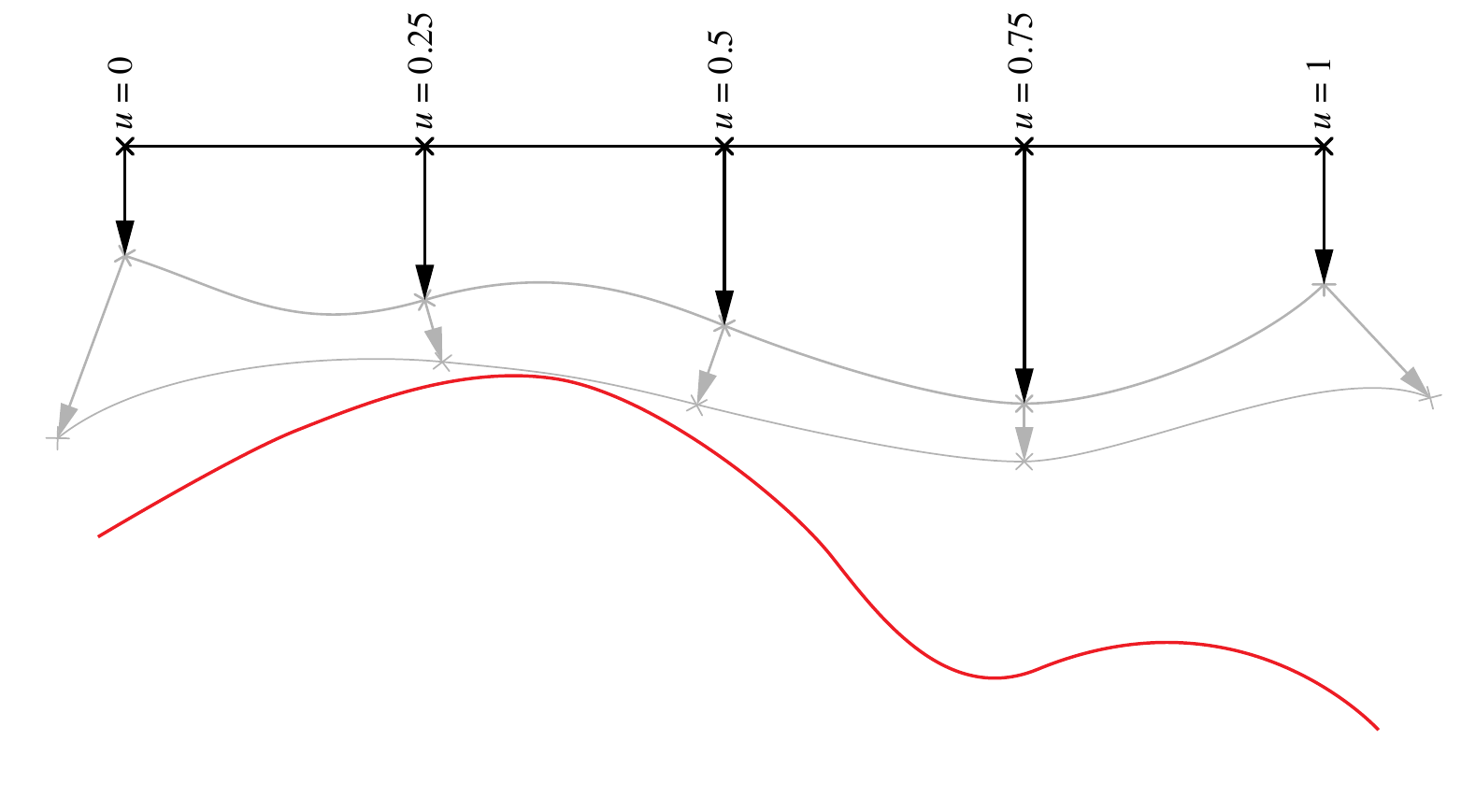}
\caption{Schematic diagrams showing the effect of a Newton line search step under different gauge fixing schemes. In both, the horizontal black line represents the initial guess surface, while the red curve represents the solution surface. \textit{Top}: When a particular source function $H_\alpha$ is specified in \eqref{eq:final2}, one is fixing both the red surface as the solution and the location of the grid points along the solution surface. The direction taken by each line search step (black arrows) consists of both the correction to the current surface (pink component) and the correction to the gauge condition (blue component). In many cases, this simultaneous gauge correction causes the nonlinear solver to become unstable. \textit{Bottom}: By setting to zero the residual of the gauge fixing equation, one no longer specifies any target gauge on the solution surface. Instead, the gauge modes are projected out from the Jacobian, leaving the line search direction with no pure gauge component. Since the gauge condition is nonlinear, one is actually making a slight change of gauge after each Newton step and therefore have no control over the gauge of the final solution. }
\label{fig:ahfinder}
\end{figure}

In the present paper, $\mathcal H$ is determined by a curve, and hence \eqref{eq:final1}-\eqref{eq:final2} become
particularly simple. We use the Newton line search to solve \eqref{eq:final1}-\eqref{eq:final2}. Note that in each
Newton step the gauge condition, \eqref{eq:final2}, is corrected. When the geometry of $\mathcal H$ becomes very
extreme, we found that the gauge condition becomes significantly more stiff than the equation for the expansion,
\eqref{eq:final1}, and the non-linear solver requires strongly suppressed step sizes or, in the worst case, fails to
converge entirely. However, the Newton solver becomes significantly more robust if we project out the gauge modes from the
line search direction entirely. To do this, instead of specifying $H(u)$ in \eqref{eq:final2} \textit{a priori}, we set
$H(u)$ to be equal to the left hand side of \eqref{eq:final2} in the current iteration. In other words, we fix the gauge
to be whatever gauge the current iteration happens to be in. Fig. \ref{fig:ahfinder} illustrates the difference between the two approaches.
In practice, we can easily implement this by fixing the residual of the gauge equation to be zero always, while still
using the Jacobian of the full system \eqref{eq:final1}-\eqref{eq:final2}. Note that with this second approach we do not
have control over the gauge of the final surface. Nevertheless, in practice we found that the if the grid points are
evenly distributed across the initial guess surface, then the solver tends to converge to the final surface in a
sensible gauge.

\begin{figure}
\flushright
\begin{overpic}[width=0.97\columnwidth]{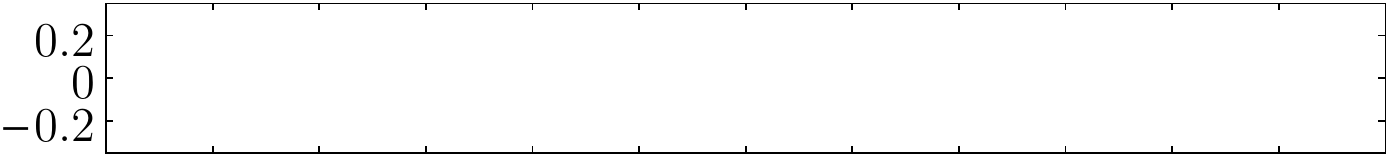}
\put(-2,0){ \includegraphics[width=0.97\columnwidth]{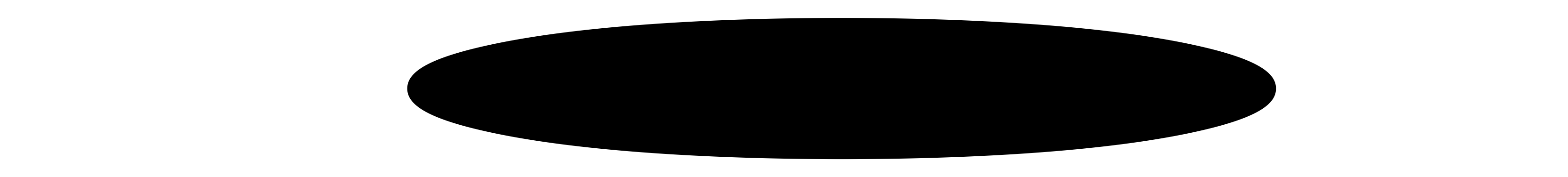}}
\put(22,16){$\hat t=5.011$}
\end{overpic}
\begin{overpic}[width=0.97\columnwidth]{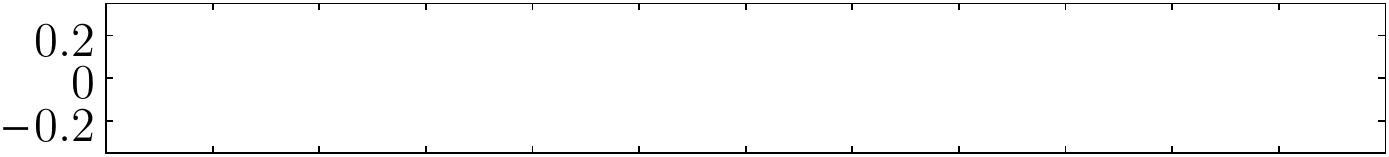}
\put(-2,0){ \includegraphics[width=0.97\columnwidth]{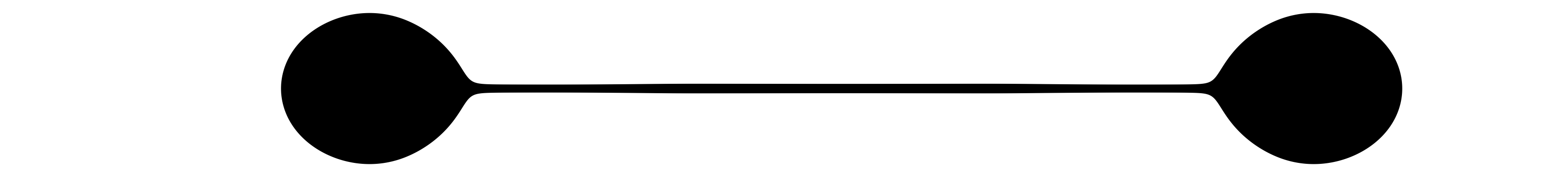}}
\put(107,17){$\hat t=26.5055$}
\end{overpic}
\begin{overpic}[width=0.97\columnwidth]{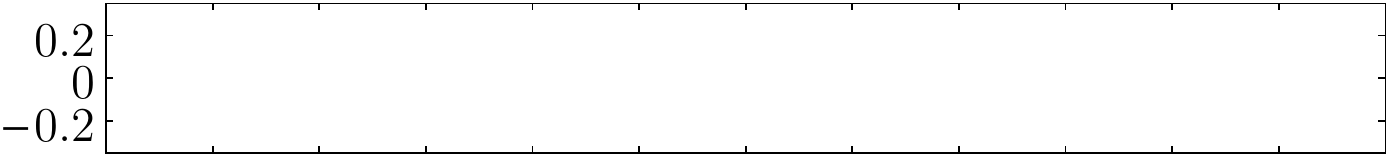}
\put(-2,0){ \includegraphics[width=0.97\columnwidth]{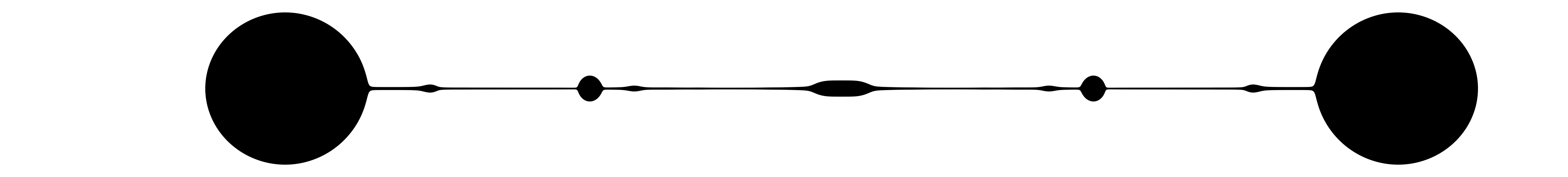}}
\put(-7,26){\rotatebox{90}{$Z$}}
\put(107,17){$\hat t=28.022$}
\end{overpic}
\begin{overpic}[width=0.97\columnwidth]{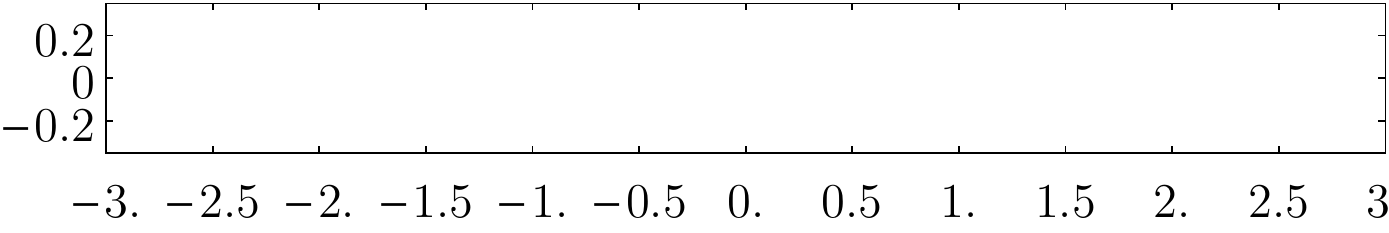}
\put(-2,0){ \includegraphics[width=0.97\columnwidth]{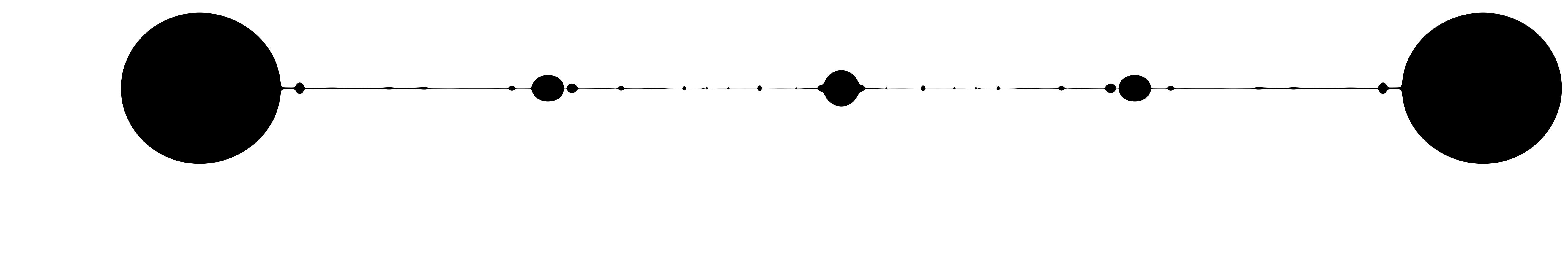}}
\put(107,30){$\hat t=28.3379$}
\end{overpic}
\par
\vspace{0.1cm}
\begin{overpic}[width=0.97\columnwidth]{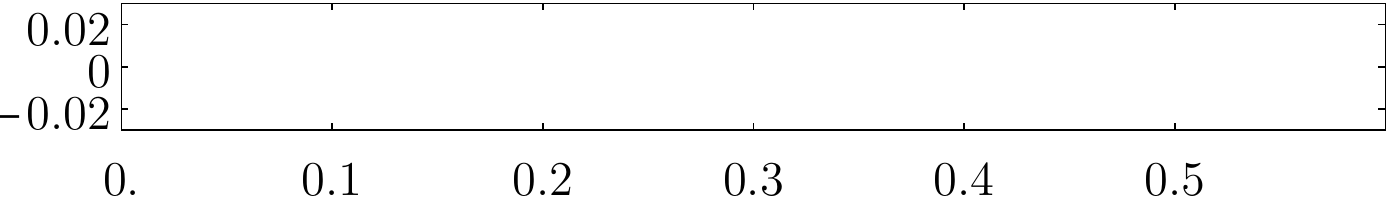}
\put(-2.4,0.35){ \includegraphics[width=0.97\columnwidth]{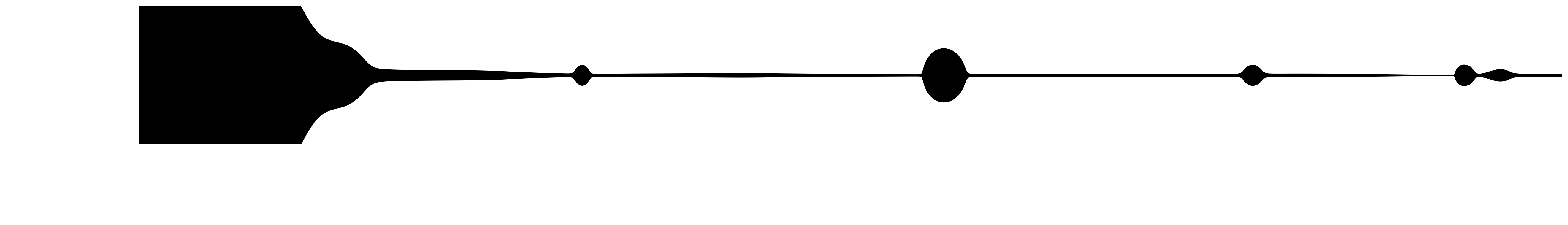}}
\put(110,-4){$U$}
\end{overpic}
\caption{Embedding diagrams of the apparent horizon at different stages of the evolution of the
ultraspinning instability of a MP BH with $a/\mu^\frac{1}{3} = 1.8$. Here $\hat t = t/\mu^\frac{1}{3}$.}
\label{fig:evolutionAH1p8}
\end{figure}

\begin{figure}
\begin{center}
\includegraphics[width=\columnwidth]{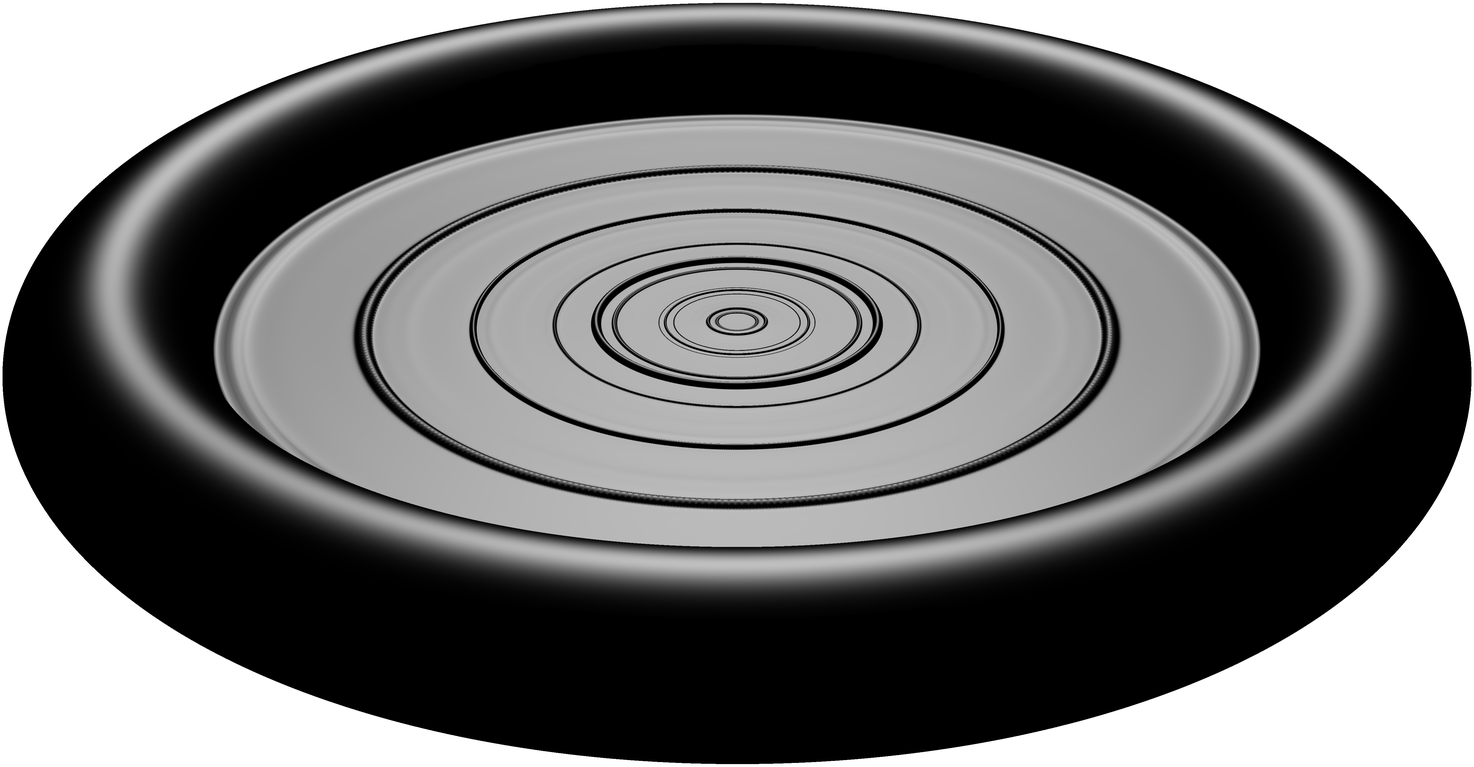}
\end{center}
\caption{Embedding of the AH for the $a/\mu^\frac{1}{3}=1.7$ simulation at time $\hat t = 36.9535$, where $\hat t = t/\mu^\frac{1}{3}$.
Five generations of concentric rings connected by black membrane sections have formed.
Almost all of the mass and angular momentum is carried by the outermost ring.}
\label{fig:3dembeds}
\end{figure}

 To visualize the apparent horizon, it is useful to consider the embedding in Euclidean space. In this Letter, we used
the same types of embeddings as \cite{Elvang:2006dd,Emparan:2014pra}. Here we consider a $\phi = \textrm{const.}$
section of the apparent horizon geometry, where $\phi$ is the rotational $U(1)$ direction, and embed it into four-dimensional Euclidean space, $\mathbf{E}_4$,
\begin{equation} \label{eq:embed}
ds^2_{\mathbf{E}_4} = dU^2 + dZ^2 + Z^2\,d\Omega_{(2)}^2\,.
\end{equation}
In Fig. \ref{fig:evolutionAH1p8} we display a sequence of snapshots of these embeddings
with the transverse sphere suppressed for the $a/\mu^\frac{1}{3}=1.8$ simulation.
In Fig. \ref{fig:3dembeds} we display a representative embedding of a $\phi = \textrm{const.}$ slice of the apparent
horizon, shown as a surface of revolution.
Note that the $\phi$-direction is not included in the embedding as it is not possible to embed the horizons of the ultraspinning
Myers-Perry black holes we consider into $\mathbf{E}_5$.
The situation is similar for the Kerr black hole, whose horizon cannot be embedded into $\mathbf{E}_3$ for high spins \cite{Smarr:1973zz}.


\begin{figure}[h]
\centering
\includegraphics[width=0.9\columnwidth]{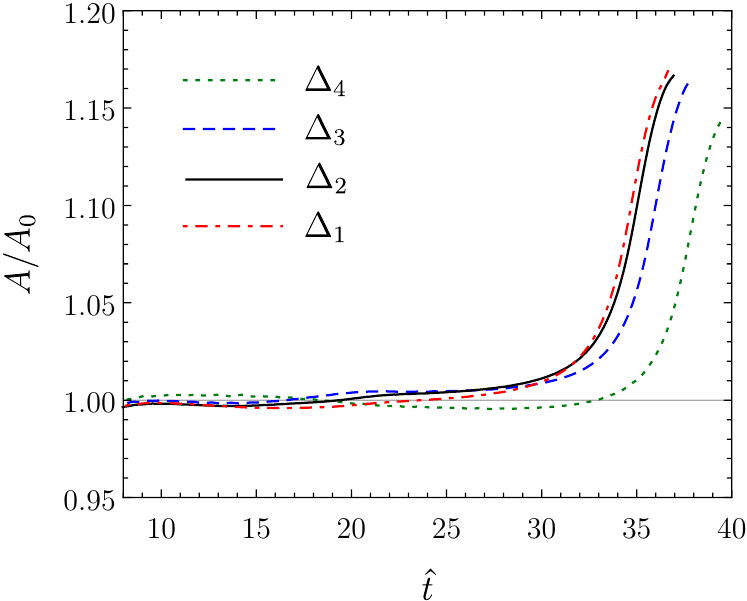}
\caption{Convergence test for the apparent horizon area.
The highest resolution run had a resolution of $\Delta_1 = 0.25 \mu^\frac{1}{3}$ on the coarsest level. For the other
runs the resolution was lowered by factors of $\sqrt{2}$. For the runs presented in the main paper we use resolution
$\Delta_2$ (solid black curve).
}
\label{fig:areaConvergence}
\end{figure}

\PRLsection{Numerical tests}
To test convergence, we produce the output presented in the main paper at four different grid resolutions. The highest
resolution run had a coarsest grid spacing of $0.25 \mu^\frac{1}{3}$. During the evolution, levels with refinement ratio
$1:2$ were added to ensure that the apparent horizon was always covered by a minimum of $80$ points.
For the runs presented in the main paper we had to add up to 22 levels.
For the lower resolutions we increase the grid spacing by factors of
$\sqrt{2}$ and correspondingly decrease the minimum number of points across the apparent horizon.
Fig. \ref{fig:areaConvergence} shows the comparison of the area of the apparent horizon from each resolution. Since the area is sensitive
to the overall structure of the horizon, this gives an indication of the accuracy level at which we can determine the properties of the
outermost ring. The results clearly show convergence; even the lowest resolution run is in the convergent
regime already. However, since the errors for the lowest two resolutions are still rather high, all results in the main paper
were obtained at the second highest resolution (solid curve in the plot).

\begin{figure}[ht]
\centering
\includegraphics[width=0.9\columnwidth]{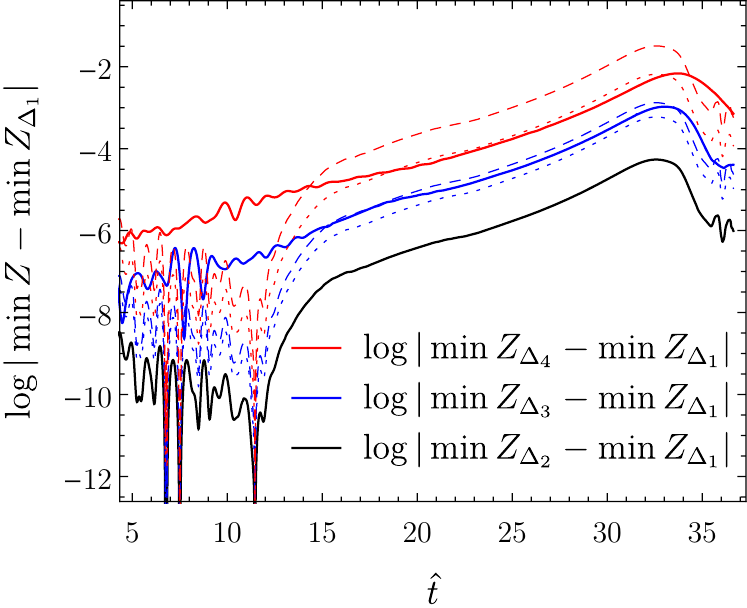}
\caption{Convergence test for the global minimum membrane thickness. Dotted and dashed lines correspond to third and
fourth order convergence respectively.
The resolutions $\Delta_{i}$ are the same as in Fig. \ref{fig:areaConvergence}.
}
\label{fig:convergence}
\end{figure}

\begin{figure}[ht]
\vspace*{0.3cm}
\centering
\includegraphics[width=0.9\columnwidth]{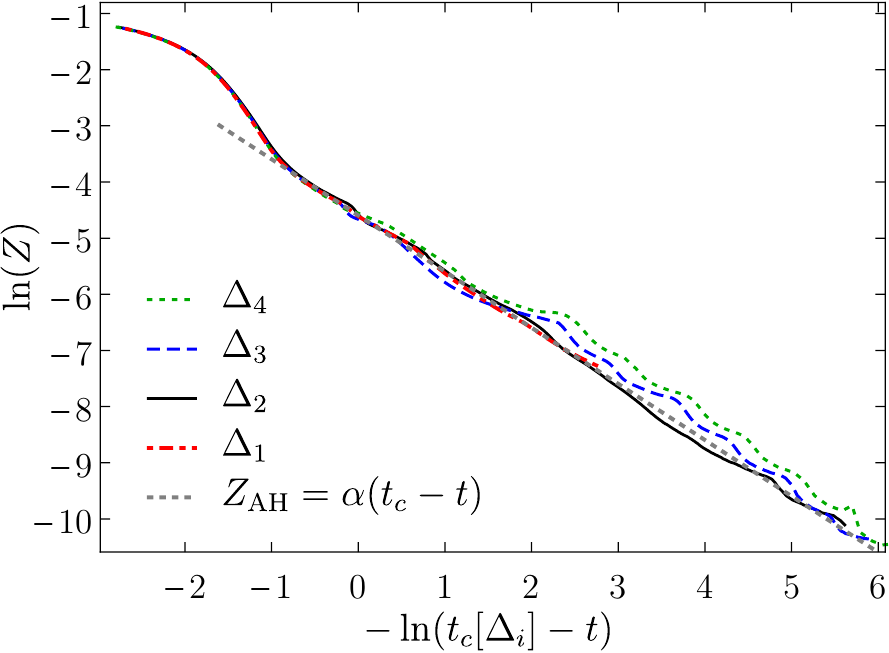}
\caption{Convergence test for the results presented in Fig. \ref{fig:thicknessPlot} in the main paper.
The resolutions $\Delta_{i}$ are the same as in Fig. \ref{fig:areaConvergence}.
The highest resolution run had to be terminated earlier than the others as it became unfeasible.
Note that the pinch-off time, $t_c$, varies slightly with the resolution (it converges with 3rd order).
To be able to compare the approach to pinch-off, each run was plotted with its specific value for $t_c$.
}
\label{fig:scalingLawConvergence}
\end{figure}

Fig. \ref{fig:convergence} shows a convergence plot for the minimum thickness of the membrane. Before the
first minimum appears, we simply plot the thickness in the middle as the middle becomes the first minimum.
This is representative of the accuracy with which we can track the growth rate of the instability and the subsequent
evolution of the membrane.
The results converge at a rate between $3^\text{rd}$ and $4^\text{th}$ order throughout the whole evolution.
This is consistent with the fact that we use a fourth order scheme but that the order is reduced due to the
interpolation at mesh boundaries.

One of our key findings is that the global minimum thickness very closely follows the scaling law
$Z_{AH} = \alpha (t_c-t)$ (see Fig.  \ref{fig:thicknessPlot} in the main paper). Fig. \ref{fig:scalingLawConvergence}
presents the same plot for four different resolutions. Our results not only converge but
also follow the scaling law increasingly more precisely as the resolution is increased.

It is important to stress that, despite the convergence results presented above, not all features in the simulation are
convergent. In particular, while we find convergence for properties of the entire black hole, the first generation and
the minimum membrane thickness, the position of higher generation rings does not converge.
Most prominently, the appearance of a central bulge is not a robust feature,
but depends on the initial data, the perturbation, and the grid setup. This is intuitive as, starting
from the second generation, the membrane is always thin enough to fit many unstable modes. Indeed, exactly the same
behavior was observed for black strings in \cite{Lehner:2010pn,Lehner:2011wc}. In the dispersion relation of the black
string, two different modes have the same growth rate, and the membrane sections arising in our simulations behave
similarly.

\end{document}